\documentstyle[12pt]{article}
\newlength{\mytopmargin}
\newlength{\myleftmargin}
\begin{document}

\vspace{4cm}
\noindent
{\large
{\bf CORRELATIONS IN TWO-COMPONENT LOG-GAS \\SYSTEMS }}
\vspace{5mm}

\noindent
A.Alastuey\footnote{email: alastuey@physique.ens-lyon.fr}\\
\noindent
Ecole Normale Sup\'{e}rieure de Lyon,Laboratoire de Physique,
Unit\'{e} de Recherche Associ\'{e}e 1325 au Centre National
de la Recherche Scientifique,
46,all\'{e}e d'Italie,69364 Lyon Cedex 07,France
\vspace{5mm}

\noindent
P.J.Forrester\footnote{email: matpjf@maths.mu.oz.au}\\
\noindent
Department of Mathematics, University of Melbourne,Parkville,Victoria
3052,Australia

\small
\begin{quote}
A systematic study of the properties of particle and  charge correlation
functions in the two-dimensional Coulomb gas confined to a one-dimensional
domain is undertaken. Two versions of this system are considered: one in
which the positive and negative charges are constrained to alternate in sign
along the line, and the other where there is no charge ordering constraint.
Both systems undergo a zero-density Kosterlitz-Thouless type transition as
the dimensionless coupling $\Gamma := q^2 / kT$ is varied through $\Gamma =
2$. In the charge ordered system we use a perturbation technique to
establish an $O(1/r^4)$ decay of the two-body correlations in the high
temperature limit. For $\Gamma \rightarrow 2^+$, the low-fugacity expansion
of the asymptotic
charge-charge correlation can be resummed to all orders in the fugacity.
The resummation leads to the Kosterlitz renormalization equations. In the
system without charge ordering the two-body correlations exhibit an $O(1/
r^2)$ decay in the high temperature limit, with a universal amplitude for
the charge-charge correlation which is associated with the state being
conductive. Low fugacity expansions establish an $O(1 /r^\Gamma)$ decay of
the two body correlations for $2 < \Gamma < 4$, and an $O(1 / r^4)$ decay
for $\Gamma > 4$. For both systems we derive sum rules which relate the long
wavelength behaviour of the Fourier transform of the charge correlations
to the dipole carried by the screening cloud surrounding two opposite
internal charges. These sum rules are checked for specific solvable
models. Our predictions for the Kosterlitz-Thouless transition and the
large-distance behaviours of the correlations should be valid at low
densities. At higher densities, both systems might undergo a first-order
liquid-gas transition analogous to the two-dimensional case.
\end{quote}


\pagebreak
\normalsize
\vspace{1.5cm}
\noindent
{\bf 1.INTRODUCTION}
\vspace{5mm}

\noindent
The natural domain for a system of charges interacting via the
two-dimensional (logarithmic) Coulomb potential is a plane. Notwithstanding
this fact, there  is still interest in studying the statistical mechanics of
log-potential Coulomb systems confined to a line (we will refer to such
systems as log-gases).
Two-component log-gases have attracted interest because of some equivalences
with
models in solid state physics (Kondo problem$^{(1)}$ and
 quantum Brownian motion problem$^{(2)}$
), and the fact that for low density a
Kosterlitz-Thouless-type pairing transition takes place as the temperature
is lowered below the critical coupling $\Gamma = 2$$^{(1,3)}$.

In this paper we will study properties of correlations in the two-component
log-gas
with oppositely signed charges of strength $q$, and a variant of this system
in which the positive and negative charges are confined to alternate in
position along the line. For couplings $\Gamma := q^2/kT \ge 1$ we also
assume that the particles are at the centre of hard rods of length $\sigma$
, which prevents short distance collapse. So as to put our work in context,
let us briefly
 review known properties of the critical behaviour and
correlations in two-component log-gas systems, and contrast this
with what is known about the two-component two-dimensional Coulomb gas
(2dCG) where
appropriate.

\vspace{.5cm}
\noindent
{\bf 1.1 The charge ordered system}

\noindent
The system with charge ordering was first studied in the context of its
application to the Kondo problem$^{(1)}$. In this seminal work, Anderson et al.
transformed the grand partition function of the system with a hard rod length
$\sigma + d\sigma$, fugacity $\zeta$ and coupling $\Gamma$ into the grand
partition function of the system with hard rod length $\sigma$ and modified
$\zeta$ and $\Gamma$, thereby deriving a pair of coupled renormalization
equations. The transformation, which is approximate, requires
$\zeta$ and $\Gamma - 2$ to be small  and is thus
applicable in the neighbourhood of the Kosterlitz-Thouless transition.
Remarkably, applying this procedure to study the 2dCG
, Kosterlitz$^{(4)}$ found essentially the same equations.
 Furthermore, for the 2dCG it is well known (see e.g. ref.[5
]) that a renormalization procedure can be applied to study the
charge-charge correlation and a certain length dependent dielectric
constant. Again, by an appropriate choice of variables, the resulting
equations are precisely those found by Anderson et al..

One immediate prediction from the renormalization
equations is the dependence of the critical coupling ($\Gamma_c$) on
fugacity: thus for the charge ordered system
$$
\Gamma_c - 2 = 2^{5/2}\zeta
\eqno (1.1)
$$
valid to leading order in $\Gamma_c - 2$ and $\zeta$.
This phase boundary separates a low-temperature dipole phase in which the
positive and negative charges are paired, from a high temperature  phase in
which the positive and negative charges are dissociated. More quantitatively,
from the mapping to the Kondo problem, Schotte and Schotte$^{(6)}$ have argued
that for the finite charge ordered system of length $L$ with periodic
boundary conditions
$$
\langle ( \sum q_i x_i - L/2 )^2 \rangle
\mathop{\sim}\limits_{L \rightarrow \infty}
\cases{c(\Gamma)L &for $\Gamma < 2 $\cr
c(\Gamma)L^2&for $\Gamma > 2$\cr}
\eqno (1.2)
$$
This behaviour, in which the boundary conditions play an essential role, was
confirmed by Monte Carlo simulation$^{(6)}$.

In contrast to the phase indicator (1.2), the phases of the 2dCG can be
distinguished by the dielectric constant $\epsilon$, which is finite in the
dipole phase and infinite in the high temperature (conductive) phase. We
recall that if the system has dielectric constant $\epsilon$, then a
fraction $1 - 1/\epsilon$ of an infinitesimal external charge density will
be screened. Furthermore, $\epsilon$ can be expressed in terms of the
particle correlations by the formulas$^{(7)}$
$$
{1 \over \epsilon} = 1 + {\pi \beta \over 2} \int d\vec{r} \, r^2 C(r)
\eqno (1.3a)
$$
and
$$
{1 \over \epsilon} = -{\beta q \over 2 \rho}
\int d\vec{r} \, \vec{F}(\vec{r})  \vec{p}_{-+}(\vec{r}) \rho_{-+}(r)
\eqno (1.3b)
$$
where $C(r)$ denotes the charge-charge correlation, $\vec{F}(\vec{r})$ denotes
the
force corresponding to the two body  potential between two positive charges
of unit strength (which is assumed to be smoothly regularized at the origin),
$\rho_{-+}(r)$ denotes the distribution function between two opposite
charges and $\vec{p}_{-+}(r)$ denotes the dipole moment of the charge
distribution induced by two fixed opposite charges separated a
distance $r$.

In log-gas systems it can be argued that a dielectric phase has $\epsilon =
1$ ( see e.g. ref.[8]). A linear response argument then gives
$$
C(r) \sim o \left ( {1 \over r^2} \right ) \qquad {\rm as } \:\:\: r
\rightarrow
\infty
\eqno (1.4a)
$$
On the other hand, in a conductive phase the charge-charge correlation must
exhibit the universal asymptotic behaviour$^{(9)}$
$$
C(r) \sim -{q^2 \over \Gamma (\pi r)^2}
\eqno (1.4b)
$$
In the sense of screening an infinitesimal external charge, the charge
ordered system is expected to always exhibit a dielectric phase (except
possibly at the phase boundary)$^{(8)}$. Hence, from (1.4a), the decay of
$C(r)$
should be faster than $1 / r^2$ in both the low temperature dipole and high
temperature
phase. When the positions of the particles are restricted to the sites of
a lattice (this mimics a hard-core regularization, the
isotherm $\Gamma = 1$ is exactly solvable$^{(9)}$. Both the truncated
two-particle
distribution functions between like and opposite charges, and thus $C(r)$,
exhibit an
asymptotic $O(1 / r^4)$ decay in accordance with this prediction.

\vspace{.5cm}
\noindent
{\bf 1.2 No charge ordering}

\noindent
Application of the rescaling method of Anderson et. al. to the two-component
log-gas without charge ordering leads to the conclusion$^{(3)}$ that the
critical coupling is $\Gamma = 2$ independent of the fugacity, in contrast
to the behaviour (1.1). No other information is obtained.

The charge-charge correlation function in the low temperature dipole phase
must again exhibit the behaviour (1.4a). This has been explicitly verified
for the solvable isotherm $\Gamma = 4$, where $O(1/ r^4)$ decay was
found$^{(10)}$. The high temperature phase is expected to be conductive and
thus $C(r)$ should obey the sum rule (1.4b). The phase boundary $\Gamma =
2$, when the positions of the particles are restricted to the sites of a
lattice, is a
solvable isotherm$^{(10,11)}$ and it is found that $C(r)$ has a $O(1 / r^2)$
decay, but with a density dependent amplitude.

\vspace{.5cm}
\noindent
{\bf 1.3 Paper outline}

\noindent
{}From the above brief review, it is clear that there are many gaps in our
knowledge of the critical properties and the behaviour of correlations in
two-component log-gas systems. To improve on this situation, we will make a
fairly systematic study of the correlation functions in the high and low
temperature phases as well as in the vicinity of the zero-density critical
point. The
charge ordered system is considered in section 2, while the system without
charge ordering is considered in section 3. Concluding remarks are made in
section 4.

In subsection 2.1 the two-particle correlations in the high temperature
phase of the charge ordered system are analysed using a perturbative
approach. The correlations in the scaling region are analysed in subsection
2.2 by applying the low density resummation method of Alastuey and
Cornu$^{(7)}$
and in subsection 2.3 the correlations in the low temperature phase away
from criticality are considered.
In subsection 2.4 the second BGY equation is transformed into Fourier space
and a sum rule analogous to (1.3b) is obtained.

In subsection 3.1, the two-particle correlations in the conductive phase of
the system without charge ordering are studied using a
linear response argument and macroscopic electrostatics, and a high
temperature resummation technique. The asymptotic
behaviour of the correlations in the dielectric phase are determined in
subsection 3.2 by studying the low fugacity expansions at $O(\zeta^4)$,
and an analysis of the second BGY equation similar to that given in
subsection 2.3 for the charge ordered model is made in subsection 3.3. From
the latter analysis the behaviour of the three and four particle
correlations in the dielectric phase is deduced.

In Section 4 we summarize our results with emphasis placed on the mechanisms
behind the contrasting behaviours of the correlations in the log-gas
with and without charge ordering.

\vspace{1cm}
\noindent
{\bf 2. THE CHARGE ORDERED LOG-GAS}

\vspace{.5cm}
\noindent
{\bf 2.1 Decay of correlations in the high temperature phase}

\noindent
Along the high temperature solvable isotherm $\Gamma = 1$ it has been shown
by explicit calculation$^{(9)}$ that the charge-charge correlation decays as
$O(1 / r^4)$, and thus by (1.4a) the system does not screen an infinitesimal
external charge. Indeed, because of the charge ordering constraint, it was
argued$^{(9)}$ that this latter property is a general feature of the system
for all temperatures. In particular, the high temperature phase is not
conductive, which suggests that the conventional methods of analysing the
high temperature phase in Coulomb systems (Debye-Huckel type theories etc.)
are not applicable.
In fact the familiar Abe-Meeron$^{(12)}$  diagrammatics, with Debye-like
mean-field potentials resulting from chain resummations, cannot be applied
here because the constraint of charge ordering is equivalent to introducing
a many-body potential between the charges, whereas the diagrammatics hold
for systems with two-body forces only.
We use instead a perturbative method which relies on the
special screening properties of the $\Gamma = 0$, constrained two-species
perfect gas reference system.

\vspace{.3cm}
\noindent
{\bf 2.1.1 Correlations at $\Gamma = 0$}

\noindent
For $\Gamma < 1$ in general, and $\Gamma = 0$ in particular, there is no
need to regularize the short range singularity of the logarithmic potential
as the corresponding Boltzmann factors are integrable. Thus at $\Gamma = 0$
the log-gas system can be considered as a two-species perfect gas of point
particles, constrained so that the two-species alternate in position along
the line. Let us suppose the first particle belongs to species $+$ while the
second particle belongs to species $-$ (these boundary conditions will
result in a symmetry breaking: the $\rho_{+-}$ and $\rho_{-+}$ correlation
functions will not in general be equal), and there are $N$ particles of each
species with coordinates $x_1, \dots ,x_N$ ($+$ species) and
$y_1, \dots ,y_N$ ($-$ species) where $x_j, y_j\in [-L/2,L/2] (j=1,\dots,N)$.
The canonical partition function $Z_{2N}$ for this system is defined as
$$
Z_{2N} :=  \left ({1 \over N!} \right )^2\int^* dx_1 \dots dy_N
\eqno (2.1a)
$$
where * denotes the interlacing constraint on the interval $[-L/2,L/2]$.
It is easily evaluated to give
$$
Z_{2N} = {1 \over (2N)!} L^{2N}
\eqno (2.1b)
$$
The two particle distribution $\rho_{++}(x_1,x_2)$ is given by
$$
\rho_{++}(x_1,x_2) := {N(N-1) \over Q_{2N}} \int^{*(x_1,x_2)} dx_3 \dots
dy_N
\eqno (2.2a)
$$
where
$$
Q_{2N} = (N!)^2Z_{2N}
\eqno (2.2b)
$$
and $*(x_1,x_2)$ denotes the interlacing constraint, given that there are
particles of species $+$ at $x_1$ and $x_2$. The distribution
$\rho_{-+}(x_1,y_1)$ is defined similarly.

To calculate $\rho_{++}(x_1,x_2)$, we note that in the interval $[-L/2,x_1]$
there must be equal numbers of particles of each species, while in the
intervals $[x_1,x_2]$ and $[x_2,L/2]$ there must be one more particle of
species $-$ than particle of species $+$. Let the number of particles of
species
$+$ in each interval be $M_a,M_b$ and $M_c$ respectively (note that $M_a + M_b
+ M_c = N - 2$). Then we have
$$
\rho_{++}(x_1,x_2) = {(N!)^2 \over Z_{2N}}\sum^{N-2}_{M_a,M_b,M_c = 0 \atop
M_a + M_b + M_c = N - 2} { (x_1+L/2)^{2M_a} (x_2 - x_1)^{2M_b + 1}
(L/2 - x_2)^{2M_c + 1} \over (2M_a)! (2M_b + 1)! (2M_c + 1)!}
\eqno (2.3)
$$
Substituting (2.1b), this can be rewritten as
$$
\rho_{++}(x_1,x_2) = {2N(N-1) \over L^{2N}} S_{2N-2}(x_1+L/2,x_2-x_1,L/2-x_2)
\eqno (2.4a)
$$
where
$$
S_p(a,b,c) :=
\sum^{p/2-1}_{M_a,M_b,M_c \atop M_a+M_b+M_c=p/2-1}
{p! \over (2M_a)!(2M_b+1)!(2M_c+1)!} a^{2M_a}b^{2M_b+1}c^{2M_c+1}
\eqno (2.4b)
$$
Using the generalized binomial expansion
$$
(a+b+c)^p =
\sum^p_{p_1,p_2,p_3=0 \atop p_1+p_2+p_3=p}
{p! \over p_1! p_2! p_3!} a^{p_1} b^{p_2} c^{p_3}
\eqno (2.5)
$$
it is straightforward to derive the summation formula
$$
S_p(a,b,c) = {1 \over 4} \left [
(a+b+c)^p + (-a + b +c)^p - (a-b+c)^p - (-a -b +c)^p \right ]
\eqno (2.6)
$$
valid for $p$ even. With this result the thermodynamic limit in (2.4a) can
be taken immediately to give
$$
\rho_{++}(0,x) = \rho^2 \left [ 1 - e^{-4 \rho |x|} \right ]
\eqno (2.7)
$$
A similar calculation shows
$$
\rho_{+-}(0,x) = \rho^2 \left [ 1 + e^{-4 \rho |x|} \right ]
\eqno (2.8)
$$
or alternatively this result could be deduced from the requirement
$$
{1 \over 2} \left ( \rho_{++}(0,x) + \rho_{+-}(0,x) \right ) =
\rho^2
\eqno (2.9)
$$
which follows since the combination of two particle distributions on the
l.h.s. gives the two-particle distribution of the (unconstrained) perfect
gas.

The crucial feature of these distributions is that they exhibit perfect
screening of an internal 'charge'. Thus
$$
\int_{-\infty}^{\infty} dx \, \left ( \rho_{++} (0,x) - \rho_{+-} (0,x)
\right ) = - \rho
\eqno (2.10)
$$
This property allows the correlations in the high temperature limit to be
studied by a perturbation expansion about the $\Gamma = 0$ constrained
perfect gas reference system.

\vspace{.3cm}
\noindent
{\bf 2.1.2 Perturbation about $\Gamma = 0$}

\noindent
In any two component log-gas system, the Coulomb interaction energy $U$ can be
written
$$
U = {1 \over 2} \int dx \int dx' \left [ Q(x) Q(x') \right ]_{\rm nc}
v_c(|x - x'|)
\eqno (2.11a)
$$
where
$$
v_c(|x-x'|) := - \log |x-x'|
\eqno (2.11b)
$$
$$
Q(x) := q \sum_{j=1}^N \left ( \delta (x-x_j) - \delta (x - y_j) \right )
\eqno (2.11c)
$$
with the positive (negative) species at $x_j{} (y_j){} $ and
$$
\left [ Q(x) Q(x') \right ]_{\rm nc}
$$
denotes that the product $Q(x)Q(x')$ is formed with products over coincident
points excluded. In the charge ordered two-component system the truncated
two particle distribution between like charges is given by
$$
\rho_{++}^T(0,x_a) = {\langle \left [ N_+(0) N_+(x_a) \right ]_{\rm nc}
e^{-\beta U}\rangle_0 \over \langle e^{-\beta U} \rangle_0} -
{ \langle N_+(0) e^{-\beta U} \rangle_0
 \langle N_+(x_a) e^{-\beta U} \rangle_0
\over \langle e^{-\beta U} \rangle_0^2}
\eqno (2.12)
$$
where the subscript $0$ indicates that the averages are taken with respect
to the constrained perfect gas reference system and
$$
N_+(x_a) = \sum_{j=1}^N \delta (x-x_j)
$$

Next we expand the exponentials in (2.12) to leading order in $\beta$. We
obtain
$$
\rho_{++}^T(0,x_a) \: \sim \:
 -{\beta \over 2} \int_{-\infty}^{\infty} \int_{-\infty}^\infty dx dx' \,
\langle N_+(0)N_+(x_a)
\left [ Q(x) Q(x') \right ]_{\rm nc} \rangle_0^T v_c(|x-x'|)
\eqno (2.13)
$$
where the truncation is defined with respect to the three quantities
$0,x_a$ and $(x,x')$.$^{(13)}$ When expressed in terms of the fully truncated
Ursell
functions there are two contributions. One involves only two body Ursell
functions while the other involves the four body Ursell function. Since the
Ursell
functions in the reference system decay exponentially, the leading contribution
comes from the term involving the two body functions.
Hence (2.13) can be rewritten as
$$
\rho_{++}^T(0,x_a) \sim - \beta \int_{-\infty}^\infty \int_{-\infty}
^{\infty}dx dx' \,
\langle N_+(0) Q(x) \rangle_0 \langle N_+(x_a) Q(x') \rangle_0 v_c(|x-x'|)
\eqno (2.14)
$$
The two body averages are given by
$$
\langle N_+(0) Q(x) \rangle_0 =
q \left ( \rho_{++}^{(0)}(x) - \rho_{-+}^{(0)}(x) +\rho_0 \delta (x) \right )
= -2q \rho_0^2 e^{-4\rho_0|x|} + q \rho_0 \delta (x)
\eqno (2.15a)
$$
and
$$
\langle Q(x') N_+(x_a) \rangle_0 =
\langle Q(x'-x_a) N_+(0)\rangle_0
\eqno (2.15b)
$$

To calculate the leading large-$x_a$ behaviour of (2.14) we make the
expansion
$$
v_c(|x-x'|) = v_c(|x_a|) +
\sum_{n=1}^\infty{(x'-x-x_a)^n \over n!} {\partial^n \over \partial x^n}
v_c(|x|) {\Big |}_{x=x_a}
\eqno (2.16)
$$
Due to the perfect screening property (2.10) of the reference system, and the
fact the distribution functions in the reference system are even in their
argument, we see that the
first non-zero term which results from substituting (2.16) in (2.14) is
$n=4$. Noting
$$
{\partial^4 \over \partial x^4} v_c(|x|) =
-{\partial^4 \over \partial x^4 }\log |x| =
{3! \over x^4}
$$
we therefore have that to first order in $\Gamma$ and for large-$x_a$
$$
\rho_{++}^T(0,x_a) \sim - { 6 \Gamma {{\rho_0}}^2 \over (4 \rho_0 x_a)^4}
\eqno (2.17)
$$

The leading asymptotics of $\rho_{+-}^T(0,x_a)$ can be computed in a similar
way. In fact, since in the reference system
$$
\rho_{+-}^T(0,x) = - \rho_{++}^T(0,x)
$$
the only difference in the calculation is a minus  sign so we obtain
$$
\rho_{+-}^T(0,x) \sim  - \rho_{++}^T(0,x)
\sim  { 6 \Gamma {\rho_0}^2 \over (4 \rho_0 x_a)^4}
\eqno (2.18)
$$

Although (2.17) and (2.18) are valid to first order in $\Gamma$ only, we
expect the leading $O(1/x^4)$ decay to persist in the high temperature phase
for $\Gamma \le 1$ at least, as the exact result at $\Gamma = 1^{(10)}$
exhibits this behaviour.
Indeed we expect all terms in the $\Gamma$-expansion of the truncated two
body distributions to decay as $1/x^4$; this would follow from a
diagrammatical analysis similar to the one used by Alastuey and
Martin$^{(13)}$.
The exact result also exhibits the property (2.18)
relating the leading asymptotics of $\rho_{+-}^T(0,x)$ and $\rho_{++}^T(0,x)$,
which suggests this may also be a general property of the high temperature
phase for $\Gamma \le 1$ at least.

\vspace{.5cm}
\noindent
{\bf 2.2 Correlations in the scaling region of the low temperature dipole
phase}

\noindent
For the 2dCG it has been proved$^{(14)}$  that all the coefficients in the low
fugacity expansion of the pressure and the correlation functions are
convergent in the dipole phase ($\Gamma \ge 4$). We expect this property
to remain true in the low temperature dipole phase of the log-gas systems
($\Gamma \ge 2$). With this assumption, we can consider the asymptotic
large-$r$ behaviour of each such coefficient in the charge-charge correlation
$C(r)$ for $\Gamma \ge 2$. In particular we can consider the behaviour for
$\Gamma \rightarrow 2^+$, as the Kosterlitz-Thouless transition is
approached from the dipole phase. Motivated by the analogy between critical
features  of the alternating model and the 2dCG (in particular the occurrence
of the same renormalization equations), we do this by closely following the
strategy of Alastuey and Cornu$^{(7)}$ in their analysis of $C(r)$ in the
two-dimensional system.

For this purpose we introduce the asymptotic charge density $C_\Delta (r)$.
This is defined as the terms in the asymptotic expansion of $C(r)$, which
when replacing $C(r')$ in
$$
\Delta := 1 + { 4 \Gamma \over q^2} \int_\sigma^\infty dr' \, r' C(r')
\eqno (2.19)
$$
give the correct leading order singular behaviour of $\Delta$ for $\Gamma
\rightarrow 2^+$ at each order in $\zeta$. Note from (1.3a) that $\Delta$ is
the analogue of $1/\epsilon$ in the 2dCG. Furthermore, in the low fugacity
limit $\Delta - 1$ is proportional to the mean distance of separation
between neighbouring charges$^{(15)}$. Our objective is to calculate
$C_\Delta(r)$ and $\Delta$ in the scaling region of the low temperature
dipole phase.

\vspace{.3cm}
\noindent
{\bf 2.2.1  Low fugacity expansions}

\noindent
The low fugacity expansions of the correlations can be obtained from the low
fugacity expansion of the logarithm of the grand partition function, with each
positive (negative) charge given a position dependent fugacity
$\zeta a(x_j){} (\zeta b(y_j))$. The grand partition function is given by
$$
\Xi [a,b] = \sum_{N=0}^\infty \zeta^{2N} Z_{2N} [a,b]
\eqno (2.20)
$$
where $Z_{2N}[a,b]$ denotes the partition function for equal number $N$ of
positive and negative charges. The truncated distributions can be calculated
from (2.20) according to the formula
$$
\rho_{+ \dots + - \dots -}^{T} (r_1, \dots ,r_{n_1};s_1, \dots ,s_{n_2})
= { \delta^{n_1+n_2} \over \delta a(r_1) \dots \delta a(r_{n_1})
\delta b(s_1) \dots \delta b(s_{n_2}) }
\log \Xi [a,b] {\Big |}_{a=b=1}
\eqno (2.21)
$$

Now from (2.20)
$$
\log \Xi [a,b] =
\zeta^2 Z_2[a,b] + \zeta^4 \left ( Z_4[a,b] - {1 \over 2}(Z_2[a,b])^2 \right )
+ O( \zeta^6 )
\eqno (2.22)
$$
Using (2.22) in (2.21), assuming that the left most particle always has a
positive charge, and writing the partition functions explicitly we
obtain for the two-particle correlations
$$
\rho_{++}^T(r) = \zeta^4 \left [ \int_\sigma^{r-\sigma} dy_1
\int_{r+\sigma}^\infty dy_2 \,\left ({r(y_2 - y_1)\over
(y_2 - r)y_2(r-y_1)y_1}\right )^\Gamma
- \left ( \int_\sigma^\infty {dy \over y^\Gamma} \right )^2 \right ]
+ O(\zeta^6)
\eqno (2.23a)
$$
$$
\rho_{+-}^T(r) = {\zeta^2 \over r^\Gamma} + \zeta^4 \left [
\int_{2\sigma}^{r-\sigma} dx \int_{\sigma}^{x-\sigma} dy \,
 \left ( {(r-y) x\over r(x-y)(r-x)y}\right )^\Gamma -
 \left ( \int_\sigma ^\infty {dy \over y^\Gamma } \right )^2 \right.
$$
$$
+  2\int_{r+2\sigma}^{\infty}dy \int_{\sigma+r}^{y-\sigma}dx \,
\left \{ \left | {x(y-r) \over (y-x)y(x-r)r }  \right |^\Gamma -
\left ( {1 \over (y-x)r } \right )^\Gamma \right \}
$$
$$
- \left.{1 \over r^\Gamma} \left ( \int_{-\sigma}^{r+ 2\sigma} dy
 \int_{-\infty}^{y-\sigma} dx + \int_{r+\sigma}^\infty dy \int_{-\infty}
^{r} dx
  \right )
{1 \over (y-x)^\Gamma }\right ] + O(\zeta^6)
\eqno (2.23b)
$$
$$
\rho_{-+}^T(r) = \zeta^4 \left [ \int_{r+\sigma}^{\infty} dy
\int_{-\infty}^{-\sigma} dx \left | {(r-x)y \over r(y-x)(y-r)x} \right |^\Gamma
- \left ( \int_\sigma^\infty {dy \over y^\Gamma } \right )^2 \right ]
+ O(\zeta^6)
\eqno (2.23c)
$$
where it is assumed $r > 2\sigma$ in (2.23a) and $r > \sigma$ in (2.23b) and
(2.23c) (for $r$ less than these values the respective full distribution
functions vanish). Also, the first double integral in (2.23b) is to be
omitted if $r < 2 \sigma$.
To the same order, the low fugacity expansion of the charge-charge
correlation follows immediately from the formula
$$
C(r) = 2q^2 \rho \delta (r) + q^2[2\rho_{++}^T(r) - \rho_{+-}^T(r)
- \rho_{-+}^T(r)]
\eqno (2.24)
$$

\vspace{.3cm}
\noindent
{\bf 2.2.2 Evaluation of $C_\Delta^{(2)}(r)$ and $C_\Delta^{(4)}(r)$}

\noindent
Let us denote the term proportional to $\zeta^{2j}$ in the low fugacity
expansions of $C(r),C_\Delta (r)$ and $\Delta $ by $C^{(2j)}(r),
C^{(2j)}_\Delta (r)$ and $\Delta^{(2j)} $ respectively. From (2.24) and
(2.23) we have, for $r > \sigma$,
$$
C^{(2)}(r) = q^2 { \zeta^2 \over r^\Gamma }
\eqno (2.25)
$$
Substituting in (2.19) gives
$$
\Delta^{(2)} = 1 + {4 \Gamma \zeta^2 \sigma^{2 - \Gamma} \over \Gamma - 2}
\eqno (2.26)
$$
We note that $\Delta^{(2)}$ is singular as $\Gamma \rightarrow 2^+$ and
furthermore to leading order is independent of $\sigma$. Both features are
true of $\Delta^{(2n)}$ in general. The latter feature implies that only the
large-$r$ portion of $C^{(2n)}(r)$ contributes to the
leading order singular behaviour of $\Delta^{(2n)}$, and thus
$C^{(2n)}_\Delta(r) $ consists of terms in the asymptotic expansion of
$C^{(2n)}(r)$. With
$n=1$ there is only one term in the asymptotic expansion, which is
$C^{(2)}(r)$ itself, so trivially $C_\Delta^{(2)}(r) = C^{(2)}(r)$.

The analysis of the large-$r$ behaviour of the integrals in (2.23), which from
(2.24) form
$C^{(4)}(r)$, is done in appendix A. There it is deduced
$$
C_\Delta^{(4)}(r) = - q^2 \zeta^4 {4 \Gamma \over r^{2 \Gamma - 2} }
\left ( - {1 \over (\Gamma - 2)^2} \left [ \left ( {\sigma \over r } \right
)^{2 - \Gamma} - 1 \right ] + {1 \over \Gamma - 2} \left ( {\sigma \over r}
\right )^{2 - \Gamma} \log r \right )
\eqno (2.27)
$$
and the integral representation
$$
C_\Delta^{(4)}(r) = - q^2 \zeta^4 {1 \over r^\Gamma} \int_{r+\sigma}^{2r} dx
\int_{x+\sigma}^{2x-r} dy {\cal S}_{(0,r)}(x,y)
\eqno (2.28a)
$$
with
$$
{\cal S}_{(0,r)}(x,y) := {4 \Gamma \over (x-r)(y-x)^{\Gamma - 1} }
\eqno (2.28b)
$$
is also derived.
The r.h.s. of (2.27) has the remarkable property of having an identical
structure to the corresponding expression in the low fugacity expansion of
the charge-charge correlation in the 2dCG [ref. 7,eq.(4.13)]. Furthermore,
the integral expression (2.28) can be interpreted as the contribution
of
a mobile positive-negative dipole (positive charge at $x$, negative charge
at $y$) pair partially screening (note that in all phases the
external charges are not screened at all since $\epsilon = 1$)
the root dipole of separation $r$ (positive
charge at 0, negative charge at $r$).
 The distance of separation $|y - x|$
between charges within the mobile dipole is constrained to be less than the
closest distance between this  dipole  and the
root charges at 0 or $r$,
 and this closest distance is to be no greater than $r$. The
factor of 4 comes from the 4 equivalent ways of arranging the mobile dipole
about the root charges (the screening dipole may lie close to 0 or $r$, and
inside or outside the root dipole).
This ''nested'' dipole interpretation of
$C_\Delta^{(4)}(r)$ is analogous to that found in ref.[7] for the quantity
$C_\epsilon^{(4)}(r)$ in the 2dCG.

\vspace{.3cm}
\noindent
{\bf 2.2.3 Nested dipole chain hypothesis}

\noindent
Although we have calculated $C^{(4)}_\Delta(r)$, it may seem a formidable
task to do likewise for $C^{(2j)}_\Delta(r)$, $j \ge 3$. Indeed, an analysis
similar to appendix A does not appear to be feasible. Instead, having
identified an analogy between $C^{(4)}_\Delta(r)$ and $C^{(4)}_\epsilon(r)$
of the 2dCG, we adapt the method given in ref. [7] to calculate
$C^{(2j)}_\epsilon(r)$.

The basis of the method of ref. [7] is the hypothesis, which has its origin
in the work of Kosterlitz and Thouless$^{(16)}$, that the configurations
contributing to $C^{(2j)}_\Delta (r)$ are all nested chains of dipoles, with
the fixed dipole (positive charge at 0 and negative charge at $r$) the
largest. The screening operator (2.28b) acts between dipoles connected in a
chain. Furthermore, by including the factor of 4 in (2.28b) these chains
can all be ordered to the right of the fixed negative charge at $r$.
For example, at $O(\zeta^6)$, there are two distinct chains as given
in Fig. 1.

The contributions to $C^{(6)}(r)$ from these chains are
$$
\int_{r +\sigma }^{2r} dx_2 \int_{x_2+\sigma}^{2x_2 -r} dy_2 \, {\cal
S}_{(0,r)}(x_2,y_2)
\int_{y_2 +\sigma }^{2y_2-x_2}dx_1 \int_{x_1+\sigma}^{2x_1 -y_2} dy_1 {\cal
S}_{(x_2,y_2)}(x_1,y_1)
$$
and
$$
\left [\int_{r+ \sigma }^{2r} dx_2 \int_{x_2+\sigma}^{2x_2 - r} dy_2 \, {\cal
S}_{(0,r)}(x_2,y_2)
\right ]
\left [\int_{r+ \sigma}^{2r} dx_1 \int_{x_1+\sigma}^{2x_1 -r} dy_1 {\cal
S}_{(0,r)}(x_1,y_1)
\right ]
$$
Furthermore, the first of these contributions needs to be weighted by a factor
of
$2$ to account for the relabelling degeneracy, and this linear combination
needs to be multiplied by a factor of $-q^2\zeta^6 /(4!r^\Gamma) $.

At general order, the nested chain hypothesis gives [ref.7  ,eqs.(4.26)-(4.28)]
$$
C_\Delta^{(2n)}(r) = - {q^2 \zeta^2 \over r^\Gamma}{\zeta^{(2n-2)} \over
(2n-2)!} S_\Delta^{(2n-2)}(r)
\eqno (2.29a)
$$
where
$$
S_\Delta^{(2n-2)}(r) =
\sum_{p=1}^{n-1} {(n-1)! \over p!(n-1-p)!}
\sum_{q_\alpha \ge 0 \atop q_1 + \dots +q_p = n-1-p} {(n-1-p)! \over q_1!
\dots q_p! } I_{2q_1}(r) \dots I_{2q_p}(r)
\eqno (2.29b)
$$
with
$$
I_{2q}(r) = \int_{r +\sigma}^{2r} dx \int_{x+\sigma}^{2x-r} dy \, {\cal
S}_{(0,r)}(x,y)
S_\Delta^{(2q)}(x-y)
\eqno (2.29c)
$$
{}From the structure (2.29) it is shown in ref.[7] that the series
$$
{C_\Delta}(r) = \sum_{n=1}^\infty C_\Delta^{(2n)}(r)
\eqno (2.30)
$$
can be summed whatever the form of ${\cal S}_{(0,r)}(x,y)$, with the result
$$
{C_\Delta}(r) = -{q^2 \zeta^2 \over r^\Gamma }
\exp \left [ - {1 \over q^2} \int_{r+\sigma}^{2r} dx \int_{x+\sigma}^{2x-r}
dy {\cal S}_{(0,r)}(x,y) (y-x)^\Gamma C_\Delta(y-x) \right ]
\eqno (2.31)
$$
Since ${\cal S}_{(0,r)}(x,y)$ is explicitly given by (2.27b), we see that it is
possible
to simplify the double integral in (2.31) by an integration by parts in the
$y$-variable. This gives
$$
C_\Delta(r) = -{q^2 \zeta^2 \over r^\Gamma}
\exp \left [ -4\beta \log r \int_\sigma^r dx \, x C_\Delta(x) +
4 \beta \int_\sigma^r dx \, \log x \, xC_\Delta (x) \right ]
\eqno (2.32)
$$
This expression has the same structure with that of $C_\epsilon(r)$ for the
2dCG [ref.7,eq.(4.32)].

Introducing the length dependent version of (2.19) by
$$
\Delta (r) := 1 + {4 \Gamma \over q^2} \int_\sigma^r dr' \, r'C_\Delta (r')
\eqno (2.33)
$$
which gives the contribution to $\Delta$ from all particles of separation at
most $r$ and is analogous to the length dependent  dielectric constant defined
by
Kosterlitz and Thouless$^{(16)}$, we can obtain from (2.32) and (2.33) the
coupled differential equations
$$
{d \over d \log (r)} \Delta (r) = {4 \Gamma \over q^2} r^2 C_\Delta (r)
\eqno (2.34a)
$$
and
$$
{d \over d\log (r)} C_\Delta (r) = - \Gamma C_\Delta (r) \Delta (r)
\eqno (2.34b)
$$
Making the change of variables
$$
y(r) = {2 \over q}\left (-\Gamma C_\Delta (r) \right )^{1/2}r, \qquad
t(r) = 1 - {\Gamma \over 2} \Delta (r), \qquad l = \log r
$$
the differential equations (2.34) read
$$
{d \over dl} y(l) = y(l)t(l)
\eqno (2.35a)
$$
$$
{d \over dl} t(l) = \left ( y(l) \right )^2
\eqno (2.35b)
$$
These equations, with different meanings to $y(l)$ and $t (l)$, are
precisely those obtained by Anderson et al.$^{(1)}$.

It is simple to solve the system (2.35) for $t(l)$ in terms of $y(l)$,
where the constant of integration can be determined by putting $r=\sigma$ in
the integral equation. We find, using $r$ instead of $l$,
$$
-{4 \Gamma \over q^2} r^2 C_\Delta (r) = \left (1 - {\Gamma \over 2} \Delta (r)
\right )^2
+ 4 \Gamma \zeta^2 \sigma^{2 - \Gamma} - \left (1 - {\Gamma \over 2}\right )^2
\eqno (2.36)
$$
Now, assuming $\Gamma > 2$,  from the low fugacity expansion $C_\Delta (r)$ is
$o(1/r^2)$ as $r \rightarrow \infty$. Recalling the
definition (2.19) and taking the limit $r \rightarrow \infty$ in (2.36) we thus
have
$$
\Delta = 1 - { \Gamma - 2 \over \Gamma} \left (\left [ 1 - { \Gamma (4\zeta)^2
\sigma^{2 - \Gamma}\over (\Gamma - 2
)^2 } \right ]^{1/2}- 1 \right )
\eqno (2.37)
$$
Notice that the radius of convergence of the resummed function in (2.37)
gives the phase boundary (1.1).
This expression for $\Delta$ has an identical structure to that of
$1/\epsilon$ in the scaling region of the dipole phase of the 2dCG$^{(7)}$.

The resummation of $C_\Delta (r)$ can be performed by analysing the
differential equations (2.34). This has been done in ref. 7 using these
equations as they apply to the correlations of the 2dCG. From the
workings of ref. 7 [eqs.(4.41)-(4.50)] we find
$$
C_\Delta(r) = - q^2 \zeta^2 \left [ A_0 \left ( {\sigma \over r} \right)^{
\Gamma
\Delta } + \sum_{N=1}^\infty A_N\left ({ \sigma \over r} \right )^{\Gamma
\Delta + N(\Gamma  \Delta - 2)} \right )
\eqno (2.38a)
$$
where
$$
A_0 = \exp \left \{ - \Gamma \int_\sigma^\infty {dt \over t}
\left [  \Delta(t) - \Delta \right ] \right \}
\eqno (2.38b)
$$
and the $A_N$ can be expressed in terms of $A_0$ and the function
$$
2 (4 \zeta )^2 /(\Gamma - 2)^2 \over [1 - 2(4 \zeta)^2 / (\Gamma - 2)^2
]^{1/2}
\eqno (2.39)
$$
Thus the coupling $\Gamma$ is renormalized as the density dependent quantity
$\Gamma \Delta$, and $\Gamma \Delta$ is the power involved in the expansion
(2.38) of $C_\Delta(r)$.

\vspace{.5cm}
\noindent
{\bf 2.3 Correlations in the low temperature dipole phase away from
criticality}

\noindent
The analysis of Appendix A identifies the leading large-$r$ expansion of
$\rho_{++}^T(r)$,$\rho_{+-}^T(r)$ and $\rho_{-+}^T(r)$ at $O(\zeta^4)$
throughout the low temperature dipole phase $\Gamma > 2$. Thus from the
expansions (A6) we have
$$
\rho_{++}^{T(4)}(r) \: \sim \: \rho_{+-}^{T(4)}(r) \: \sim \:
\rho_{-+}^{T(4)}(r) \: \sim \: {\zeta^4 \Gamma \over (\Gamma - 2)^2 r^2}
\eqno (2.40)
$$
In the case of $\rho_{++}^{T(4)}(r)$ this behaviour results from the
contribution of the configuration A of figure 2 in Appendix A.
Similarly, configurations in which each root particle is paired with
a mobile particle to form a neutral cluster give the leading contribution to
$\rho_{+-}^{T(4)}(r)$ and $\rho_{-+}^{T(4)}(r)$.
These configurations have the same statistical weight at large distances,
which implies the same behaviour (2.40) for $\rho_{++}^{T(4)},
\rho_{+-}^{T(4)}$ and $\rho_{-+}^{T(4)}$.

To calculate the large-$r$ behaviour of $\rho_{++}^{T(2n)}(r)$ etc. for
$n > 2$ we need to hypothesize the generalization of configuration A which
is dominant in the respective integral formulas for $r \rightarrow
\infty$. This generalization is to have all particles belonging to a neutral
cluster about and including one or other of the root particles, with the
interparticle spacing within a cluster small in comparison to $r$. The
potential $V_{2N}$ between clusters can be written as
$$
V_{2N} = W_0+W_r+U_{0r}
\eqno (2.41)
$$
where $W_0$ and $W_r$ denote the electrostatic energies of the individual
clusters about and including the root particles at 0 and $r$ respectively,
and $U_{0r}$ denotes the mutual interaction. For large-$r$, $U_{0r}$ is
to leading order a dipole-dipole potential:
$$
U_{0r} \: \sim \: p_0 {\partial \over \partial x_0} p_r
{\partial \over \partial r} \log |x_0 - r| {\Big |}_{x_0=0}
= -{p_0p_r \over r^2}
\eqno (2.42)
$$
where $p_0$ and $p_r$ are the dipoles of the clusters about and including
the root particles. The situation for $n > 2$ is thus completely analogous
to the case $n = 2$ (recall the text below (A2)). Insertion in the
corresponding integral formulas, and use of the expansion
$$
e^{-\beta U_{0r}} \: \sim \: 1 - \beta U_{0r} + \dots
\eqno (2.43)
$$
thus leads to the results
$$
\rho_{++}^{T(2n)}(r) \: \sim \: \rho_{+-}^{T(2n)}(r) \: \sim \:
\rho_{-+}^{T(2n)}(r) \: \sim \: {\zeta^{2n} \alpha_{2n}(\Gamma)\over  r^2}
\eqno (2.44)
$$
where the coefficient $\alpha_{2n}(\Gamma)$ diverges as $\Gamma \rightarrow
2$. This $O(1/r^2)$ behaviour can be understood as resulting from the
dipole-dipole interaction between the neutral clusters at large separation.

The configurations giving the leading large-$r$ behaviour of higher order
correlations will again be neutral clusters and, analogous to
the case $n=2$, the potential between far away neutral clusters will again
be to leading order due to the dipole-dipole interaction and thus
$O(1/r^2)$. Hence we expect
$$
\rho_{+\dots + - \dots -}^{T(2n)}(x_1,\dots,x_{n_1};y_1,\dots,y_{n_2})
\: \sim \: {\zeta^{2n} \over \prod x_{jk}^2 \prod y_{j'k'}^2 \prod
|x_{j}-y_{k'}|^2}
\eqno (2.45)
$$
as $x_{jk}:=x_k-x_j, y_{j'k'}:=y_{k'}-y_{j'}, |x_{j}-y_{k'}| \rightarrow
\infty$, where we have
omitted the amplitude. This asymptotic behaviour requires at least one mobile
particle about each root particle, and thus does not necessarily apply for
$n < n_1 + n_2$.

\vspace{.5cm}
\noindent
{\bf 2.4 A sum rule from the BGY equation for C(x)}

\noindent
It has been shown in ref. 7 that the BGY equation for the charge-charge
correlation $C(r)$ in the 2dCG can be used to derive a sum rule which
expresses the dielectric constant in terms of the dipole moment $p_{+-}(r)$ of
the screening cloud surrounding two internal charges of opposite sign. In
this subsection we will apply an analogous analysis to $C(x)$ (we use $x$ to
denote any position coordinate, positive or negative and $r$ to denote a non-
negative quantity)
for the charge
ordered system which leads to a sum rule involving the distribution
functions between charges of opposite sign and the dipole moments $p_{+-}(x)$
and $p_{-+}(x)$.

The BGY equation for $C(r)$ in the 2dCG has been derived from the BGY
equations for the two-particle distributions. The BGY equation for $C(x)$ in
the two-component log-gas without charge ordering can immediately be read
off from the 2dCG result (ref.7, eq.(5.10)). However, the resulting equation
is not applicable to the charge ordered system for two reasons: (i) the
charge ordered constraint introduces extra terms in the BGY equation for the
two-particle distributions
and thus for $C(x)$, and (ii) the symmetry $\rho_{+-}(x)=\rho_{-+}(x)$
assumed in  the derivation is not valid in the low temperature phase of the
charge ordered system.

To derive the BGY equations for $C(x)$ in the charge ordered system let us
then reconsider the BGY equations for the two-particle distribution functions.
We suppose
the logarithmic potential is regularized by hard cores of length $\sigma$
symmetrically placed about each particle. We know that without charge ordering,
and with smoothly regularized potentials
$$
v_{s_1s_2}(x) = s_1s_2v(x)
$$
the BGY equation for the distribution $\rho_{s_1s_2}^T(x)$ can be written in
the
form (ref. 7, eqs. (5.8a),(5.8b))
$$
{\partial \over \partial x_2}\rho_{s_1s_2}^T(x_{12}) =
s_1s_2\Gamma F(x_{12})\rho_{s_1s_2}^T(x_{12}) +
s_2\beta q \rho^2 \int_{-\infty}^\infty dx_3 F(x_{32}) Q_{s_1}(x_1|x_3)
$$
$$
+s_2 \Gamma \int_{-\infty}^\infty dx_3 \, F(x_{32}) [ \rho_{s_1s_2+}^T(x_1,
x_2, x_3) -
\rho_{s_1s_2-}^T (x_1,x_2,x_3) ]
\eqno (2.46a)
$$
where
$$
F(x) := -{\partial \over \partial x}v(x), \qquad x_{ab} := x_b - x_a
\eqno (2.46b)
$$
and $Q_{s_1 \dots s_n}(x_1, \dots ,x_n|x)$ denotes the total charge density
induced at $x$ given that there are charges $s_1, \dots ,s_n$ fixed at
$x_1, \dots ,x_n$:
$$
Q_{s_1 \dots s_n}(x_1,\dots,x_n|x) =
q {[ \rho_{s_1 \dots s_n +}(x_1, \dots ,x_n,x) -
\rho_{s_1 \dots s_n -}(x_1, \dots, x_n,x) ] \over
\rho_{s_1 \dots s_n}(x_1, \dots ,x_n) }
$$
$$
+ q \sum_{i = 1}^n s_i \delta (x - x_i)
\eqno (2.46c)
$$
By considering the definition of
$\rho_{s_1s_2}(x)$ in the canonical ensemble, it is easy to see that the only
modifications needed to this equation due to the charge ordering constraint
and the hard cores are that
$$
\rho_{s_1s_2(-s_2)}(x_1,x_2,x_2-\sigma) -
\rho_{s_1s_2(-s_2)}(x_1,x_2,x_2+\sigma)
\eqno (2.47)
$$
must be added to the r.h.s., and the smoothed potential $v(x)$  replaced by
$v_c(x)$ (recall (2.11b)).
The definition of $Q_{s_1}(x_1|x_3)$ remains that given by (2.46c)
and the integration over $x_3$ remains extended over the whole real line. This
is a
consistent prescription as the r.h.s. of (2.46a) can then be expressed entirely
in
terms of full distribution functions which vanish for configurations with
$|x_i-x_j|<\sigma$.
However, the equation (2.46a) modified by the hard core terms (2.47) is now
valid for $|x_{12}|> \sigma$ (notice that the l.h.s. of (2.46a) diverges at
$|x_{12}| = \sigma$)

Adding (2.47) to the r.h.s. of (2.46a), multiplying both sides by
$q^2s_1s_2$, summing over $s_1,s_2=\pm$ and use of the definitions (2.24)
and (2.46c) gives that the BGY equation for $C(x)$ in the charge ordered system
with hard cores is
$$
{\partial \over \partial x_2} \left ( C(x_{12}) -2q^2 \rho \delta (x_{12})
\right )
 = 2\Gamma \rho \int_{-\infty}^\infty dx_3 F_c(x_{32})C(x_{13}) +
\Gamma q \int_{-\infty}^\infty dx_3 \, F_c(x_{32})R(x_1,x_2,x_3)
$$
$$
+ qA(x_1,x_2)
\eqno (2.48a)
$$
where
$$
R(x_1,x_2,x_3) = \sum_{s_1,s_2,s_3=\pm}qs_1s_3\rho_{s_1s_2s_3}^T(x_1,x_2,x_3)
+ q \sum_{s_1s_2}\rho_{s_1s_2}^T(x_1,x_2)\delta(x_3-x_1)
\eqno (2.48b)
$$
$$
\doteq \sum_{s_2,s_3 =\pm } s_3
 \rho_{s_2s_3}(x_2,x_3)Q_{s_2 s_3}(x_2,x_3|x_1)
-{2 \rho \over q}C(x_{31})
\eqno (2.48b')
$$
(the symbol $\doteq$ denotes that (2.48b$'$) gives the same value as (2.48b)
when substituted in (2.48a); see the sentence below (2.48d))
and
$$
A(x_1,x_2) = B(x_1,x_2,x_2-\sigma) - B(x_1,x_2,x_2 + \sigma)
\eqno (2.48c)
$$
with
$$
B(x_1,x_2,x) =  \rho_{+-}(x_2,x) Q_{+-}(x_2,x|x_1)
- \rho_{-+}(x_2,x) Q_{-+}(x_2,x|x_1)
$$
$$
-q{\Big (} \delta (x_1 - x_2) - \delta (x_1 -x) {\Big )}
{\Big (} \rho_{+-}(x_2,x) + \rho_{-+}(x_2,x) {\Big )}
\eqno (2.48d)
$$
In deriving (2.48) we have used the equation
$$
\int_{-\infty}^{\infty}dx \, F_c(x) \sum_{s_2,s_3=\pm} s_2s_3\rho_{s_2s_3}(0,x)
= 0
$$
which is a consequence of the first factor in the integrand being odd while
the second factor is even.

In preparation for analysing (2.48a) we note that on the l.h.s we can make
the replacement
$$
{\partial \over \partial x_2} \mapsto -{\partial \over \partial x_1}
\eqno (2.49)
$$
without changing its value.
Similarly, on the r.h.s. we can write
$$
F_c(x_{32}) = -{\partial \over \partial x_2} v_c(x_{23})
=  {\partial \over \partial x_3} v_c(x_{32})
\eqno (2.50)
$$
The coordinate $x_2$ is now not involved in any operation, so for convenience
it can be set equal to zero. Let us now proceed to transform the modified
BGY equation into Fourier space.
To do this we multiply both sides of the $x_2$-independent form of
 (2.48a) by $e^{ikx_1}$ and integrate over $x_1$ with the condition $|x_1|
 > \sigma$.
On the l.h.s. we have
$$
\left ( \int_{-\infty}^{-\sigma} + \int_\sigma^\infty \right )
dx_1 \, e^{ikx_1} \left ( - { \partial
\over \partial x_1 } [ C(x_1) - 2q^2 \rho \delta (x_1) ] \right ),
\eqno (2.51a)
$$
which after integration by parts becomes
$$
-2i \sin (k\sigma) C(\sigma) + ik[\tilde{C}(k) - 2q^2 \rho]
\eqno (2.51b)
$$

On the r.h.s. we can extend the integration to the entire real line because
this side vanishes for $|x_1| < \sigma$ and remains finite at
$|x_1| = \sigma$.
 Now it follows from (2.48b) that $R(x_1,x_2,x_3)$  is
 a fully truncated quantity and thus the integral involving $R$
on the r.h.s. of (2.48a) is absolutely convergent. The integral over $x_1$
of this term can therefore be done before the integral over $x_3$.
Doing this and using the convolution theorem in the first term gives
that the Fourier transform of the r.h.s. equals
$$
- 2 \Gamma \rho i k {\tilde v_c}(k) {\tilde C}(k)
+\Gamma q \int_{-\infty}^{\infty} dx_3 {\partial v_c(x_3) \over \partial x_3}
{\tilde R}(k;x_3)
+  q {\tilde A}(k)
\eqno (2.52a)
$$
where, from (2.48b$'$) and (2.48c)
$$
{\tilde R}(k;x_3) = \sum _{s_2,s_3 = \pm} s_3 \rho_{s_2s_3}(0,x_3){\tilde Q}_
{s_2s_3}(0,x_3|k) -({2 \rho \over q})e^{ikx_3}\tilde{C}(k)
\eqno (2.52b)
$$
and
$$
q{\tilde A}(k)  =
 2iq^2 \sin(k\sigma) \Big ( \rho_{+-} (\sigma) + \rho_{-+}
(\sigma ) \Big )
 +q\rho_{-+} (\sigma)\left (
{\tilde Q}_{+-} (0,-\sigma | k) + {\tilde Q}_{-+}(0,\sigma |k) \right )
$$
$$
-q\rho_{+-} (\sigma) \left (
{\tilde Q}_{+-} (0,\sigma |0) + {\tilde Q}_{-+}(0,-\sigma |x_1) \right )
\eqno (2.52c)
$$
(in deriving (2.52c) we have used the symmetry $\rho_{+-}(x) = \rho_{-+}
(-x)$, where $\rho_{s_1s_2}(x):=\rho_{s_1s_2}(0,x)$).

To obtain the desired sum rule, we consider the leading order small-$k$
behaviour of (2.51) and (2.52). The term independent of $k$ on both sides
vanishes due to sum rules for the perfect screening of an internal charge:
$$
{\tilde C}(0) = \int_{-\infty}^{\infty}dx \, C(x) = 0
\eqno (2.53a)
$$
and
$$
{\tilde Q}_{s_a s_b} (x_a,x_b;0) :=
\int_{-\infty}^{\infty} dx Q_{s_as_b}(x_a,x_b|x) = 0
\eqno (2.53b)
$$
which are expected to be true in all phases of Coulomb systems$^{(17)}$.
The leading small-$k$ term is thus proportional to $k$. From (2.51), on the
l.h.s. it is given by
$$
-2 i k \sigma C(\sigma) - 2 i k q^2 \rho
\eqno (2.54a)
$$
while from (2.52) the leading small-$k$ terms on the r.h.s. are
$$
-2\Gamma \rho i k \tilde{v_c}(k) \tilde{C}(k)
+ik\Gamma q
\sum_{s_2,s_3 = \pm} s_3 \int_{-\infty}^\infty dx_3 \,
{ \partial v_c(x_3) \over \partial x_3 } \rho_{s_2s_3}(0,x_3)p_{s_2s_3}(0,x_3)
$$
$$
+ 2 i q^2 k \sigma [ \rho_{+-} (\sigma) + \rho_{-+} (\sigma) ]
+ i k q \rho_{-+}(\sigma) [ p_{+-}(0,-\sigma) + p_{-+}(0,\sigma) ]
$$
$$
- i k q \rho_{+-}(\sigma) [ p_{+-}(0,\sigma) + p_{-+}(0,-\sigma) ]
\eqno (2.54b)
$$
where
$$
p_{s_1s_2}(0,x_3) := \int_{-\infty}^\infty dx \, xQ_{s_1s_2}(0,x_3|x)
\eqno (2.54c)
$$

Let us now equate (2.54a) and (2.54b). The resulting equation can be
simplified by recalling that $\rho_{++}(x) = 0$ for all $|x| < 2 \sigma$,
and so
$$
C(\sigma) = -q^2 \left ( \rho_{+-}(\sigma) + \rho_{-+} (\sigma) \right )
$$
Furthermore, ${\tilde v_c}(k) =  \pi /|k|$, and by symmetry
$ p_{++}(x) = p_{--}(x) = 0 $ for all $x$ in any phase.
We thus obtain the sum rule
\begin{eqnarray*}
\lefteqn{\lim_{k \rightarrow 0} \pi \beta {\tilde C}(k) /|k|} \\
& &= 1 + {\beta q \over 2 \rho} \int dx \, {\partial v_c \over \partial x}
\left ( \rho_{-+} (x)p_{-+}(x) - \rho_{+-}(x)p_{+-}(x) \right )
\end{eqnarray*}
$$
+ {1 \over q \rho} \left ( \rho_{-+}(\sigma)p_{-+}(\sigma) -
\rho_{+-}(\sigma)p_{+-}(\sigma) \right )
\eqno (2.55)
$$
which is to be obeyed in all phases. However we have commented in subsection
1.1
that for the charge ordered system $C(r)$ is expected to exhibit a decay which
is
$o(1/r^2)$, except possibly on the phase boundary. Such a decay implies
${\tilde C}(k)=o(|k|)$ and thus the l.h.s. of (2.55) vanishes, leaving us
with a simplified form of the sum rule which is to be obeyed by the system in
all phases except
possibly on the phase boundary. Note in particular that the dipole moment
$p_{+-}(x)$ is non-zero in both phases. In Ref. [9] it was claimed that
$p_{+-}(x)$ as calculated from the exact results at $\Gamma =1$ vanished.
However, in reviewing that calculation, an error has been found and the correct
conclusion is that $p_{+-}(x)$ is non-zero at $\Gamma = 1$, in agreement
with the general property.

\vspace{.3cm}
\noindent
{\bf 2.4.1 Verification of the sum rule for the $\Gamma = 0$ reference system}

\noindent
We have shown in subsection 2.1 that the $\Gamma = 0 $ reference system
perfectly screens an internal charge. Since the perfect screening
properties (2.53) are the two fundamental assumptions in deriving the
sum rule (2.55) from the BGY equation (2.48a), we therefore expect the sum rule
to be satisfied in the reference system. Let us check this property.

The reference system of subsection 2.1 consists only of point
particles, so the hard core width in (2.55) needs to be taken to zero. Also,
the
final terms on the r.h.s can be simplified due to the symmetries $\rho_{+-}
(x) = \rho_{-+}(x),\, p_{+-}(x) = -p_{-+}(x)$, and the pair potential in the
reference system is identically zero. In these circumstances the sum rule
(2.55) reads
$$
0 = 1 - {2 \over q \rho} \rho_{+-}(0)p_{+-}(0^+)
\eqno (2.56)
$$
($p_{+-}(x)$ is discontinuous at the origin so it is necessary to specify its
value on one side).

{}From (2.9), $\rho_{+-}(0)=2 \rho^2$. It remains to
calculate
$p_{-+}(x)$. From the definitions (2.54c) and (2.46c) we must first calculate
$\rho_{+-+}^{(3)}(x_1,x_2,x_3)$ and $\rho_{+--}^{(3)}(x_1,x_2,x_3)$,
which can be done by an appropriate extension of the method given in
subsection 2.1 to calculate $\rho_{+-}^{(2)}(x)$. We find
$$
\rho_{+-+}(x_1,x_2,x_3) = \cases{\rho^3(1 + e^{- 4 \rho |x_3 - x_2|}
+ e^{-4 \rho |x_2 - x_1|} + e^{-4 \rho |x_3 -x_1|} ),&for $x_1 < x_2 <x_3$\cr
\rho^3(1 + e^{- 4 \rho |x_3 - x_2|}
- e^{-4 \rho |x_2 - x_1|} - e^{-4 \rho |x_3 -x_1|} ),&for $x_1<x_3<x_2$\cr}
\eqno (2.57)
$$
and
$$
\rho_{+--}(x_1,x_2,x_3) = \cases{\rho^3(1 - e^{-4 \rho|x_3 - x_2|}
+ e^{-4 \rho|x_2 -x_1|} - e^{-4 \rho |x_3 - x_1|} ),&for $x_1<x_2<x_3$\cr
\rho^3(1 + e^{-4 \rho|x_3 - x_2|}
+ e^{-4 \rho|x_2 -x_1|} + e^{-4 \rho |x_3 - x_1|} ),&for $x_2<x_1<x_3$\cr}
\eqno (2.58)
$$

Using (2.57), (2.58) and (2.9) to form $Q_{+-}(x_1,x_2|x)$ as given by (2.46c)
we can
check the perfect screening property
$$
\int_{-\infty}^\infty dx \, Q_{+-}(x_1,x_2|x) = 0
\eqno (2.59)
$$
which together with (2.53a) is a fundamental assumption in the derivation of
(2.53). We can also calculate the first moment of $Q_{+-}$ to obtain
$$
\rho_{+-}(x)p_{+-}(x) = \pm {q \rho \over 2} e^{-4 \rho |x|}
\eqno (2.60)
$$
where the positive sign is to be taken for $x>0$, while the negative sign
holds for $x < 0$.

{}From (2.60)
we see that indeed the sum rule (2.55) is obeyed for the
$\Gamma = 0$ reference system.

\vspace{1cm}
\noindent
{\bf 3. THE SYSTEM WITHOUT CHARGE ORDERING}

\vspace{.5cm}
\noindent
{\bf 3.1 Decay of correlations in the conductive phase}

\noindent
For $\Gamma < 2$ the two-component log-gas without charge ordering is
expected to perfectly screen an infinitesimal charge density. As noted in
subsection 1.2, a linear response argument$^{(8)}$ then leads to the sum rule
(1.4b) for the asymptotic behaviour of the charge-charge correlation.
In this subsection we will note another derivation of (1.4b), using a new
method due to Jancovici$^{(18)}$, which is based on a linear response argument
and
macroscopic electrostatics.
In ref. [18] the conductive phase is characterised by the applicability of
macroscopic electrostatics at large length scales in the system. The log-gas
in a conductive phase is thus considered  as an infinite metal line obeying
the laws of two-dimensional electrostatics. When combined with a linear
response relation, describing the change in the potential due to the
addition of an external charge, this characterisation implies (1.4b).
Furthermore it is clear that (1.4b) only applies to the leading
non-oscillatory term of the charge-charge correlation only: the use of
macroscopic electrostatics implies the charge density must be smoothed
over some microscopic distance and thus the oscillatory terms averaged to
zero.

\vspace{.2cm}
\noindent
{\bf 3.1.1 Abe-Meeron type resummations}

\noindent
In three-dimensional classical Coulomb systems a systematic way to study the
high-temperature conductive phase is via the Abe-Meeron$^{(12)}$
diagrammatic expansion of the Ursell function $h_{s_1s_2}(r)$ ($:=
\rho_{s_1 s_2}^T(r)/\rho^2$). This approach is also applicable to
the present system.

Briefly, let
us recall (see ref. [19] for a detailed recent review of the method, and an
extension to quantum systems) that $h_{s_1 s_2}(r)$ is given by the
sum of all the Mayer graphs built with the Mayer bonds
$$
f_{s_1s_2}(r) = e^{- \Gamma v_{s_1 s_2}}(r) - 1.
$$
(here the potential $ v_{s_1 s_2}$ corresponds to the regularized
logarithmic Coulomb
potential between unit charges of signs $s_1$ and $s_2$).
Once the chain summations have been performed, the above set of Mayer graphs
is exactly transformed into an a new set of `prototype' graphs $\Pi$ with
the same topological structure and two kinds of resummed bonds:\\ (i) the
Debye-like bond $\phi_{s_1 s_2}(r)$ defined as the sum of all the
convolution chains built with the Coulomb potential $s_1 s_2
v_c(r)$, and such that
$$
\tilde{\phi}_{s_1 s_2}(k) = - {\Gamma s_1 s_2 \tilde{v}_c(k) \over
1 + 2 \Gamma \rho \tilde{v}_c(k)};
$$
(convolutions of this bond are  to be excluded).\\
(ii) the resummed bond $f_R$:
$$
f_R(r) = e^{- \Gamma (v_{s_1 s_2}(r) - s_1 s_2 v_c(r))
+  \phi_{s_1s_2}}(r)
-1 - \phi_{s_1s_2}(r)
$$

 At large-distances, $v_{s_1 s_2}(r) - s_1 s_2 v_c(r)$ vanishes and so
 $f_R(r)$ behaves as $\phi_{s_1 s_2}^2(r)/2$. Now, since
 $\tilde{v}_c(k) \sim \pi / |k|$ for small $|k|$ we deduce from the definition
 of the Debye-like bond that $\phi_{s_1 s_2}(r)$ decays as $1 /r^2$ for
 large-$r$. Therefore all the resummed bonds (i) decay at least as
 $1/r^2$ while the resummed bonds (ii) decay as $1/r^4$; the Ursell
 functions $h_{s_1s_2}(r)$ therefore exhibit a $1/r^2$ decay in the
 conductive phase. This is to be contrasted to the $1/r^4$ decay found
 in the high temperature phase of the system with charge ordering, where
 the Abe-Meeron diagrammatics do not apply due to the presence of
 a many body
 potential inducing the charge ordering constraint.

\vspace{.5cm}
\noindent
{\bf 3.2 Decay of correlations in the dipole phase}

\noindent
Our objective in this subsection is to use low fugacity expansions to deduce
the large-$r$ decay of the truncated two-particle distributions and the
charge-charge correlation
for $\Gamma \ge 2$. Since in the system without charge
ordering $\rho_{++}(r) = \rho_{--}(r)$ and $\rho_{+-}(r) = \rho_{-+}(r)$ it
suffices to consider $\rho_{++}^T(r)$ and $\rho_{+-}^T(r)$.
We will study in detail the terms of $O(\zeta^4)$ since, as
is argued below, the same asymptotic decay is expected of terms  of higher
order in $\zeta$.

At order $\zeta^4$ the low fugacity expansions of the two particle
correlations can be calculated from (2.21), (2.22) and the explicit forms of
the partition functions which occur therein. We find
$$
\rho_{++}^{T(4)}(r) = \zeta^4 \int^* dx_1 dx_2 \left [ {1 \over 2}
\left | {r (x_2 - x_1 ) \over (x_2 - r)(x_1 - r)x_1x_2 }\right |^\Gamma
- {\rm S} { 1 \over |x_1|^\Gamma |r-x_2|^\Gamma } \right ]
\eqno (3.1)
$$
and \\[.15cm]
${}\qquad \rho_{+-}^{T(4)}(r) = $
$$
\zeta^4 \int^* dx_1 dx_2 \left [
{\rm S} \left |{ x_1(x_2 - r) \over r(x_2 - x_1)(x_1 - r)x_2} \right
|^\Gamma - {1 \over r^\Gamma |x_1 - x_2 |^\Gamma }
- {\rm S} {1 \over |x_2|^\Gamma |x_1 - r|^\Gamma} \right ]
\eqno (3.2)
$$
where
$$
{\rm S}f(x_1,x_2) := {1 \over 2} \left [ f(x_1, x_2) + f(x_2,x_1) \right ]
\eqno (3.3a)
$$
and the notation $*$ denotes the region of integration
$$
\{(x_1,x_2) \in {\bf {\rm R}}^2: |x_1-x_2| >  \sigma, \: |x_1|,|x_2|>
\sigma, \:
|x_1-r|,|x_2-r|>
\sigma\}
\eqno (3.3b)
$$
{}From these expansions, and (2.24), we have\\[.2cm]
${}\qquad
C^{(4)}(r) =
$
$$
q^2\zeta^4 \int^* dx_1 dx_2 \left [
\left | {r (x_2 - x_1 ) \over (x_2 - r)(x_1 - r)x_1x_2 }\right |^\Gamma -2
{\rm S} \left |{ x_1(x_2 - r) \over r(x_2 - x_1)(x_1 - r)x_2 }\right |^\Gamma
+ {2 \over r^\Gamma |x_1 -x_2|^\Gamma } \right ]
\eqno (3.4)
$$

In Appendix B a method of analysis of the large-$r$ behaviour of the
integral (3.4) will be given. We find that for $2<\Gamma <4$ the region of the
integrand which gives the leading order contribution comes from the physical
configurations $\alpha$ of Figure 5 in Appendix B, while for $\Gamma >4$ the
configurations $\beta$ of Figure 5 in Appendix B give the leading order
behaviour. Furthermore this remains true of the integrals (3.1) and (3.2).
Here we will calculate the contribution to the large-$r$ behaviour of the
integrals in (3.1) and (3.2) from these configurations.

For the configurations $\alpha$ of Fig. 5, which give the leading large-$r$
behaviour for $2 < \Gamma < 4$, we can replace $|x_1-r|^\Gamma$
and $|x_2-r|^\Gamma$ by $r^\Gamma$ in (3.1) and (3.2), provided we also
multiply
by a factor of 2 due to the invariance of the integrand under the mappings
$x_2-r \mapsto x_2, x_1-r \mapsto x_1$. This gives
$$
\rho_{++}^{T(4)}(r) \: \sim \: {2\zeta^4 \sigma^2 \over (\sigma r)^\Gamma}
\int^* dx_1 dx_2 \left [{1 \over 2} {|x_2 - x_1|^\Gamma \over
|x_1|^\Gamma |x_2|^\Gamma } - {\rm S} {1 \over |x_1|^\Gamma }\right ]
\eqno (3.5)
$$
and
$$
\rho_{+-}^{T(4)}(r) \: \sim \: {2\zeta^4 \sigma^2 \over (\sigma r)^\Gamma}
\int^* dx_1 dx_2 \left [{\rm S} {|x_1|^\Gamma \over |x_2 - x_1|^\Gamma
|x_2|^\Gamma } - {1 \over |x_1 - x_2|^\Gamma} - {\rm S} {1 \over
|x_1|^\Gamma} \right ]
\eqno (3.6)
$$
where $*$ now denotes the region of integration
$$
\{(x_1,x_2) \in {\bf {\rm R}}:|x_1 - x_2| > 1, \: |x_1|,|x_2| > 1\}
\eqno (3.7)
$$
and we have removed the $\sigma$ dependence from the integrals  by changing
variables
$x_1 \mapsto \sigma x_1, x_2 \mapsto \sigma x_2$. Note that the integrals in
(3.5) and (3.6) are conditionally convergent: for fixed $x_1$ (or $x_2$) one
must add the contributions of positive and negative $x_2's$ (or $x_1's$).

For $\Gamma > 4$, when the configurations $\beta$ of Fig. 5 give the
dominant contribution to the asymptotics, the integrals (3.1) and (3.2) are
analysed by expanding the integrands for $x_1$ near 0 and $x_2$ near $r$
(again we need to multiply by a factor of 2 to account for the
symmetry in the interchange
of $x_1$ and $x_2$). This gives
$$
\rho_{++}^{T(4)}(r) \: \sim \: \int^* {du dv \over |uv|^\Gamma }
\left [ \Gamma A(u,v;r) + {\Gamma^2 \over 2} \left ({uv \over r^2} \right )
^2 \right ]
\eqno (3.8)
$$
and
$$
\rho_{+-}^{T(4)}(r) \: \sim \: \int^*{du dv \over |uv|^\Gamma }
\left [ -\Gamma A(u,v;r) + {\Gamma^2 \over 2} \left ({uv \over r^2} \right )
^2 \right ]
\eqno (3.9)
$$
where
$$
A(u,v;r) := {uv \over r^2} + {1 \over 3} \left [ \left ({u-v \over r} \right
)^3 + \left ( {v \over r} \right )^3 - \left ( {u \over r} \right )^3 \right
] - { 1 \over 4} \left [ \left ( {u-v \over r} \right )^4 -
\left ( {v \over r} \right )^4 - \left ( {u \over r} \right )^4 \right ]
\eqno (3.10a)
$$
and $*$ denotes the integration region
$$
\{(u,v)\in {\bf {\rm R}}^2: \: |u| > \sigma, |v| > \sigma \}
\eqno (3.10b)
$$
In performing the integrations, only those terms in the integrand even in
$u$ and $v$ survive, so (3.8) and (3.9) simplify to give
$$
\rho_{++}^{T(4)}(r) \: \sim \: \zeta^4 { 2 \Gamma ( \Gamma - 3) \over
r^4 } \left ( { 1 \over ( \Gamma - 3 ) \sigma^{\Gamma - 3}} \right )^2
\eqno (3.11)
$$
and
$$
\rho_{+-}^{T(4)}(r) \: \sim \: \zeta^4 { 2 \Gamma ( \Gamma + 3) \over
r^4 } \left ( { 1 \over ( \Gamma - 3 ) \sigma^{\Gamma - 3}} \right )^2
\eqno (3.12)
$$

The formulas (3.8) and (3.9) have some similarities and some differences in
relation to the analogous results for the 2dCG (ref. 7 eq. (3.17)). One
similarity is that the $O(1/r^4)$ terms proportional to $\Gamma^2$ result
from squaring the dipole-dipole potential $uv/r^2$ between the neutral
clusters which form the dominant configurations. Another similarity is that
$A(u,v;r)$  is the multipole expansion (appropriately
truncated) of the potential between the neutral clusters, and that all terms
odd in the mappings $v \mapsto -v$ or $u \mapsto -u$ vanish. A crucial
difference is that in one space dimension the log-potential is not harmonic,
so unlike the two-dimensional case, the integration over $A(u,v;r)$ does not
vanish. The first non-vanishing term is the quadrupole-quadrupole potential,
which has the same $O(1/r^4)$ decay as the square of the dipole-dipole
interaction. This term gives a contribution of the same magnitude to both
(3.8) and (3.9) but of opposite sign.

At the coupling $\Gamma = 4$ the leading order behaviour of
$\rho_{++}^{T(4)}(r)$ is given by the sum of (3.5) and (3.11), and the leading
order behaviour of $\rho_{+-}^{T(4)}(r)$ is given by the sum of (3.6)
and (3.12).
At the coupling $\Gamma = 2$ all regions of integration analysed in Appendix
B were bounded by an $O(1/r^2)$ decay.
However, it is possible to  evaluate
the integrals in (3.1) and (3.2) directly by using partial fractions. This is
done below,  where it is shown
$$
\rho_{++}^{T(4)}(r) \: \sim \: {2\pi^2 \zeta^4 \over 3r^2}
\eqno (3.13)
$$
and
$$
\rho_{+-}^{T(4)}(r) \: \sim \: -{2\pi^2 \zeta^4 \over 3r^2}
\eqno (3.14)
$$
In fact these results are precisely the values given by (3.5) and (3.6) at
$\Gamma = 2$; the coefficient of the $1/r^\Gamma$ term is therefore
continuous as $\Gamma \rightarrow 2^+$.

For general $n > 2$ the large-$r$ behaviour of $\rho_{++}^{T(2n)}(r)$ and
$\rho_{+-}^{T(2n)}(r)$ can be deduced by hypothesizing the generalization of
the configurations $\alpha$ and $\beta$ of Fig. 5, which give the leading
large-$
r$ contribution, for $2 < \Gamma <4$ and $\Gamma > 4$ respectively, to the
integral expression for $\rho_{++}^{T(4)}(r)$ and $\rho_{+-}^{T(4)}(r)$. The
obvious generalization of configurations $\alpha$, which should give  the
leading large-$r$ expansion for $2 < \Gamma <4$, have $n - 2-p$ mobile
particles about one root particle and the remaining $n+p$, with
$0 \le p \le n-2$ mobile particles
about the other root particle. The mobile particles are to be
distributed so that the total charge of the cluster about and including one
root particle is $+q$ while the total charge about and including the other
root particle is $-q$. At large distance the effective potential is thus
$q^2 \log r$ which gives the required behaviour as
$$
\rho_{++}^{T(2n)}(r) \: \sim \: { A_{++}^{(2n)}(\Gamma) \over r^\Gamma }
\zeta^{2n}
\eqno (3.15)
$$
and
$$
\rho_{+-}^{T(2n)}(r) \: \sim \: { A_{+-}^{(2n)}(\Gamma) \over r^\Gamma }
\zeta^{2n}
\eqno (3.16)
$$
This is the same order decay as exhibited by (3.5) and (3.6) in the case $n=2$,
and
furthermore the amplitudes could, in principal, be given integral
representations analogous to those for $n=2$.

In particular, the coefficients in (3.15) and (3.16) should be finite as
$\Gamma \rightarrow 2^+$. This would imply that $C(r)$ as calculated from
a resummation of (3.15) and (3.16)
would exhibit the same power law decay $O(1/r^\Gamma)$, independent of the
fugacity $\zeta$. We therefore expect that the critical line separating the
conductive and insulator phases to be at $\Gamma =2$, independent of the
fugacity, which is contrary to what we have found for the charge ordered
system, but in agreement with previous studies$^{(3,20)}$. Thus the nested
dipoles of subsection 2.2.3 do not give the
dominant contribution to $C(r)$ in the system without charge ordering.
In fact the analysis in Appendix B shows that the leading contribution of
nested dipoles cancels out when the charge ordering constraint is removed.

The generalizations of the configurations $\beta$ of Fig. 5, which should give
the leading behaviour for $\Gamma > 4$, are neutral clusters about and
including the root particles. The potential $V_{2N}$ between these neutral
clusters can be expanded  as
$$
V_{2N} = W_0 + W_r + U_{0r}
$$
where $W_0$ and $W_r$ are the electrostatic energies of the neutral clusters
about 0 and $r$ respectively. For large-$r$, the potential $U_{0r}$
can be written as a multipole expansion in $r$. Analogous to the case
$n=1$, there will be two classes of terms contributing to the final leading
order
expansions
$$
\rho_{++}^{T(2n)}(r) \: \sim \: { \zeta^{2n} A_{++}^{(2n)}(\Gamma) \over r^4 }
\eqno (3.17)
$$
and
$$
\rho_{+-}^{T(2n)}(r) \: \sim \: { \zeta^{2n} A_{+-}^{(2n)}(\Gamma) \over r^4 }
\eqno (3.18)
$$

\vspace{.5cm}

\noindent
{\bf 3.2.1 Decay at $\Gamma = 2$}

\noindent
Here we will show how (3.13) and (3.14) can be derived. Consider for
definiteness
(3.13). From the Cauchy identity
$$
\left ( {r(x_2 - x_1) \over (x_2 - r)(x_1 - r) x_1 x_2} \right )^2 =
\left ({\rm det} \left [ {1/x_1 \atop 1/x_2} \quad {1/(x_1 - r) \atop
1/(x_2 - r)} \right ] \right )^2
$$
we see that at $\Gamma = 2$ (3.1) can be simplified to read
$$
\rho_{++}^{T(4)}(r) = - \zeta^4 \int^* dx_1 dx_2
{1 \over (x_2 - r)(x_1 - r)x_1 x_2}
\eqno (3.19)
$$
where the integration domain $*$ is that specified by (3.3b). The integration
over $x_2$ can now be performed using the simple partial fraction expansion
$$
{1 \over (x_2 - r)x_2 } = { 1 \over r} \left ( {1 \over x_2 - r} -
{1 \over x_2} \right )
\eqno (3.20)
$$
We must consider separately the regions $-\infty < x_1 < -2\sigma$,
$-2 \sigma < x_1 < \sigma$, $\sigma < x_1 < 2\sigma$ and $2\sigma < x_1
< 1/2$ (for $x_1 > 1/2$ an identical contribution to (3.19) is obtained).
This gives, after changing variables $x_1 \mapsto ru$,
$$
\rho_{++}^{T(4)}(r) \sim - {2 \zeta^4 \over r^2} \left \{
\int_{- \infty}^{-2 \sigma /r} + \int_{2 \sigma /r}^{1/2}
{du \over (u-1)u} \left [ \log \left | {u + \sigma /r \over
u - \sigma /r} \right | \right ] \right . \hspace{2cm}
$$
$$
 + \int_{-2 \sigma /r}^{-\sigma /r} {du \over (u-1)u} \left [ \log
\left | {\sigma /r \over u - \sigma / r} \right | - \log \left |
{-1 + \sigma / r \over u - 1 - \sigma /r} \right | \right ]
$$
$$
\left . + \int_{\sigma /r}^{2 \sigma /r} {du \over (u-1)u}\left [ \log
\left | {u + \sigma /r \over \sigma /r} \right | - \log \left |
{u - 1 + \sigma / r \over -1 - \sigma / r} \right | \right ] \right \}
\eqno (3.21)
$$
where we have ignored an additive term $ \log [(1+\sigma /r) / (1-\sigma
/r)]$ in each integrand since it decays as $r \rightarrow \infty$. In
fact the only portion of the integrand which does not decay as  $r
\rightarrow \infty$ comes form the region $u = O (\sigma / r)$. Setting
$u = x \sigma / r$ and keeping only those terms which are non-zero gives
$$
\rho_{++}^{T(4)}(r) \sim {4 \zeta^4 \over r^2} \left ( \int_1^2 dx \,
{1 \over x} \log (1 + x) + \int_2^\infty dx \, {1 \over x} \log {x + 1
\over x - 1} \right )
\eqno (3.22)
$$
Evaluating these integrals gives the result (3.13).

\vspace{.5cm}

\noindent
{\bf 3.3 A sum rule from the BGY equation for C(x)}

\noindent
Our objective in this subsection is to derive sum rules analogous to (2.55)
for the system without charge ordering. We will consider first the case when
the logarithmic attraction between opposite charges is smoothly regularized.
In this case we will  provide an illustration of the sum rule using a
solvable "parallel line" model at $\Gamma = 2$. The other case to be
considered is the hard core regularization used in the above subsection.

Let us denote the smoothly regularized potentials by
$v_{s_1s_2}(x)$ and define the corresponding force by
$$
F_{s_1s_2}(x) = -{ d \over dx} v_{s_1s_2}(x)
\eqno (3.23)
$$
(using this notation rather than that of the subsection 2.4
 allows for the possibility of different
regularization between like and unlike species).
The BGY equation for $\rho_{s_1s_2}(x_{12})$ is then given by (2.46a) with
$s_is_jF(x_{ij})$ therein replaced by $F_{s_is_j}(x_{ij})$. Assuming the
symmetry
$$
F_{s_1s_2}(x) = F_{(-s_1)(-s_2)}(x),
$$
it is straightforward to show from the BGY equation for $\rho_{s_1s_2}(x_{12})$
that the BGY equation for the charge-charge correlation in this system is
(recall the derivation of (2.48a) from (2.46a))
\setcounter{equation}{24}
\renewcommand{\theequation}{3.\arabic{equation}}
\begin{eqnarray}
\lefteqn{ {\partial \over \partial x_2}
[ C(x_{12}) - 2 q^2 \rho \delta (x_{12}) ]} \nonumber \\
& & =  \Gamma  \rho  \int_{-\infty}^\infty dx_3 \left [ F_{++}(x_{32})
- F_{+-}(x_{32}) \right ] C(x_{13}) \nonumber \\
& & + 2 \Gamma q \int_{-\infty}^\infty dx_3 F_{++}(x_{32}) \left [
\rho_{++}(x_{32}) Q_{++}(x_2,x_3|x_1) -(\rho /2q) C(x_{31}) \right ]\nonumber
\\
& & - 2 \Gamma q \int_{-\infty}^\infty dx_3 F_{+-}(x_{32})
\left [ \rho_{+-}(x_{32}) Q_{-+}(x_2,x_3|x_1) -(\rho /2q) C(x_{31}) \right ]
\end{eqnarray}

We now follow the procedure detailed in subsection 2.4 to study the
small-$k$ behaviour of the Fourier transformed version of (3.24). The final
result is the sum rule
$$
\lim_{k \rightarrow 0} \pi \beta {\tilde C}(k) /|k| =
1 - {\beta q  \over \rho} \int_{-\infty}^\infty dx \,
{\partial v_{+-}(x) \over \partial x} \rho_{-+}(x) p_{-+}(x)
\eqno (3.25)
$$
where the dipole moment $p_{-+}(x)$ is defined by (2.54c). In the distinct
phases different terms in (3.25) have special values which allows further
simplification. In the conductor phase $\Gamma < 2$ we expect $p_{-+}(x) =
0$, which implies
$$
\pi \beta {\tilde C}(k) / |k| = 1
\eqno (3.26)
$$
This result is equivalent to (1.4b). On the other hand, in the insulator
phase ($\Gamma > 2$) the expected decay (1.4a) implies the l.h.s. of (3.25)
vanishes and thus the dipole moment $p_{-+}(x)$ is non-zero.

In the case of a hard core regularisation of the logarithmic potential,
working analogous to that given in subsection 2.4 for the charge ordered
system shows that the sum rule (2.55) is still applicable. Without charge
ordering the correlations obey
$$
\rho_{+-}(x) p_{+-}(x)=-\rho_{-+}(x) p_{-+}(x)
$$
 so the sum rule reads
$$
\lim_{k \rightarrow 0} \pi \beta {\tilde C}(k) / |k|
= 1 - {\beta q \over \rho} \int_{-\infty}^\infty dx \,
{\partial v_c \over \partial x} \rho_{+-}(x) p_{+-}(x)
$$
$$
-{2 \over q \rho} \rho_{+-}(\sigma) p_{+-}(\sigma)
\eqno (3.27)
$$
In the insulator phase ($\Gamma > 2$) we expect the behaviour (1.4a) so the
l.h.s. of (3.27) will vanish. In the conductor phase ($\Gamma < 2$) the dipole
moment $p_{+-}(x)$ should vanish and thus (3.26) is reclaimed.

\vspace{.3cm}
\noindent
{\bf 3.3.1 Verification of the sum rule for a parallel line model at
$\Gamma = 2$}

\noindent
The 2dCG is exactly solvable at the coupling $\Gamma = 2^{(11)}$, where the
multi-particle distribution functions can be calculated. The method of exact
calculation used in ref.[21] can be adapted to a model in which the negative
charges (coordinates $x_j, j=1,\dots,N$) are confined to a line and the
positive charges (coordinates $X_j, j=1,\dots,N$) are confined to another
line, parallel to the first line and separated by a distance $\epsilon$. The
pair potential between opposite charges of unit strength is thus given by
$$
v_{-+}(x - X) =   \log [(x-X)^2 + \epsilon^2]^{1/2}
\eqno (3.28)
$$
while the pair potential between like charges is the Coulomb logarithmic
potential. Notice that (3.28) is a smoothly regularized Coulomb potential.
As such, the system should obey the sum rule (3.25). Let us verify this fact at
the solvable coupling $\Gamma = 2$.

To verify (3.25) we need to evaluate $p_{-+}(x_1,x_2)$, which requires the
two and three particle distributions. Using the technique of ref.[21], we
can easily evaluate the general distribution function as
\setcounter{equation}{0}
\renewcommand{\theequation}{3.29\alph{equation}}
\begin{eqnarray}
\lefteqn{\rho_{+ \dots +- \dots -}(X_1, \dots X_{n_1};x_1,\dots,x_{n_2})}\\
& &= {\rm det} \left [
{[ G_{++}(X_j,X_k) ]_{j,k=1,\dots,{n_1}} \atop
[ G_{+-}(X_j,x_k) ]_{j=1,\dots,n_1 \atop k=1,\dots,n_2}}  \qquad
{[G_{-+}(x_j,X_k)]_{j=1,\dots,n_2 \atop k=1,\dots,n_1} \atop
[G_{--}(x_j,x_k)]_{j,k=1, \dots,n_2}} \right ]
\end{eqnarray}
where
\begin{equation}
G_{++}(x) = G_{--}(x) = \int_0^\infty d \gamma \,{ e^{2 \pi i x \gamma } \over
1 + (1/2 \pi \zeta )^2 e^{ 4 \pi \epsilon \gamma} }
\end{equation}
and
\begin{equation}
G_{-+}(x) = G_{+-}(x) = -{ 1 \over 2 \pi i \zeta}
\int_0^\infty d \gamma \,{e^{2 \pi \epsilon \gamma + 2 \pi ix \gamma} \over
1 + (1/2 \pi \zeta)^2 e^{4 \pi \epsilon \gamma} }
\end{equation}

We want to use these formulas to evaluate $p_{+-}(x)$ as given by (2.54c). For
this
purpose we note from (2.46c) that
$$
\rho_{-+}(x_1,x_2)Q_{-+}(x_1,x_2;x) =
q \left ( \rho_{+-+}^T(x,x_1,x_2) - \rho_{--+}^T(x,x_1,x_2) +
\alpha(x;x_1,x_2) \right )
\eqno (3.30a)
$$
where
$$
\alpha(x;x_1,x_2) = \rho \rho_{+-}^T(x,x_1) + \rho \rho_{++}^T(x,x_2)
- \rho \rho_{--}^T(x,x_1) - \rho \rho_{-+}^T(x,x_2)
$$
$$
+ \left ( \delta (x - x_2) - \delta (x -x_1) \right ) \rho_{-+}(x_1,x_2)
\eqno (3.30b)
$$
Now it follows from the perfect screening sum rule (2.53a) for a single
internal
charge (which can be verified using (3.29)), and the dependence of the
truncated
two particle distributions $\rho_{s_1,s_2}^T(x,x')$ on $|x-x'|$, that
$$
\int_{-\infty}^\infty dx \, x \alpha (x;x_1,x_2) =
q (x_2 -x_1) \rho_{-+}^T(x_1,x_2)
\eqno (3.31)
$$
Next, to evaluate
$$
\int_{-\infty}^{\infty} dx \, x
\left ( \rho_{+-+}^T(x,x_1,x_2) - \rho_{--+}^T(x,x_1,x_2) \right )
\eqno (3.32)
$$
we note from (3.29) that
\renewcommand{\theequation}{3.33}
\begin{eqnarray}\lefteqn{
\rho_{+-+}^T(x,x_1,x_2) - \rho_{--+}^T(x,x_1,x_2)} \nonumber \\
& & = 2 {\rm Re} \left ( G_{++}(x,x_2)G_{+-}(x_2,x_1)G_{-+}(x_1,x)
- G_{++}(x,x_1) G_{+-}(x_1,x_2) G_{-+}(x_2,x) \right ) \nonumber \\
\end{eqnarray}
which shows it suffices to evaluate
$$
\int_{-\infty}^\infty dx \, xG_{++}(r,x_2) G_{-+}(x_1,x)
\eqno (3.34)
$$
This latter task can be carried out with the aid of the formula
$$
\int_{-\infty}^\infty dx \, xe^{2 \pi i x (\gamma_1 - \gamma_2)}
= { 1 \over 2 \pi i} { \partial \over \partial \gamma_1 }
\delta (\gamma_1 - \gamma_2)
\eqno (3.35)
$$
and integration by parts. Once having evaluated (3.32), we substitute the
result in the integral over $r$ of (3.30a) together with (3.31) to conclude
\begin{eqnarray*}\lefteqn{
\rho_{-+}(x_1,x_2)p_{-+}(x_1,x_2) =} \\
& & {i q \epsilon \over (2 \pi \zeta)^2} \left [
\int_0^\infty d \gamma \, {e^{2 \pi \epsilon \gamma + 2 \pi i \gamma (x_2
-x_1) } \over 1 + (1 / 2 \pi \zeta )^2 e^{4 \pi \epsilon \gamma } }
\int_0^\infty d \gamma_1 \, {e^{2 \pi \epsilon \gamma_1 + 2 \pi i \gamma_1 (x_2
-x_1) } \over (1 + (1 / 2 \pi \zeta )^2 e^{4 \pi \epsilon \gamma_1})^2  }
- ( x_1 \leftrightarrow x_2 ) \right ]
\end{eqnarray*}
$$
+ q (x_2 - x_1) \rho_{-+}^T(x_1,x_2)
\eqno (3.36)
$$

With the result (3.36) and the formula
$$
{\partial v_{-+}(x) \over \partial x}= {1 \over 2} \left (
{ 1 \over x + i \epsilon} + { 1 \over x -i\epsilon }
\right )
$$
for the force between opposite charges, all quantities in the integrand on
the r.h.s. of the sum rule (3.25) are known. The integral can now be carried
out by aid of the formula
$$
\int_{-\infty}^\infty dx \,{e^{2 \pi i (\gamma_1 - \gamma_2) x} \over x +
i\epsilon } = - 2 \pi i e^{ 2 \pi (\gamma_1 - \gamma_2) \epsilon}
\times \cases{1,&$\gamma_1 - \gamma_2 < 0$\cr
0, &otherwise\cr}
$$
We find
$$
\int_{-\infty}^\infty dx \, \rho_{-+}(x) p_{-+}(x) {\partial v_{-+}(x) \over
\partial x}
= {q \rho \over 2} \left ( 1 - {1 \over 1 + (1 / 2 \pi \zeta)^2 }\right )
\eqno (3.37)
$$
which thus evaluates the r.h.s of (3.25).

To evaluate the l.h.s. of (3.25), we note from the exact formula (3.29) the
large-$x$ expansion
$$
C(x) \: \sim \: - { q^2 \over 2( \pi x)^2 } { 1 \over 1 + 1 / (2 \pi \zeta)^2}
$$
which implies
$$
\lim_{k \rightarrow 0} \pi \beta  {\tilde C}(k)/|k|
\: \sim \: { 1 \over 1 + 1 / (2 \pi \zeta)^2 }
\eqno (3.38)
$$
Substituting (3.37) and (3.38) in (3.25) we see that the sum rule is indeed
valid.

\vspace{.5cm}
\noindent
{\bf 3.4 Decay of 3 and 4 body correlations in the dipole phase}

\noindent
The decay of the higher order correlations can be studied using the low
fugacity
expansion method of subsection 3.2. Alternatively, the decay of the 3 body
correlations can be deduced from knowledge of the decay of the 2 body
correlations and use of the BGY equation. Let us consider this latter method.

Suppose for definiteness that the logarithmic potential is smoothly regularized
at the origin and $v_{s_1s_2}=s_1s_2v$, so that the BGY equation (2.46a)
applies. Making the changes
(2.49) and (2.50) allows the Fourier transform with respect to $x_1$ of both
sides of (2.46a) to be taken, with the result
\renewcommand{\theequation}{3.39}
\begin{eqnarray}
-ik\tilde{\rho}_{s_1s_2}^T(k) & = & s_1s_2 \Gamma \int_{-\infty}^\infty dx \,
{\partial v(x) \over \partial x}
 \rho_{s_1s_2}^T(x)e^{ikx} - {s_1s_2 \over 2} \Gamma \rho i k \tilde{v}
(k)\tilde{C}(k) \nonumber \\
& & + s_2 \Gamma \int_{-\infty}^\infty dx_3 \,
{\partial v(x_3) \over \partial x_3}
\left (\tilde{\rho}_{s_1s_2+}^T
(k,0,x_3) - \tilde{\rho}_{s_1s_2-}^T(k,0,x_3) \right )
\end{eqnarray}
Now from (3.5) and (3.6), for $2 < \Gamma < 4$
$$
\rho_{s_1s_2}^T(x) \mathop{\sim} \limits_{x \rightarrow \infty} {1 \over
|x|^\Gamma}
\qquad \mbox {and so} \qquad  \tilde{\rho}_{s_1s_2}^T(k)
\mathop{\sim}   \limits_{ k \rightarrow 0} |k|^{\Gamma - 1}
\eqno (3.40)
$$
(here and below the numerical amplitudes in the asymptotic formulas are
omitted).
The first term on both sides of (3.39) is thus $O(|k|^\Gamma)$ (here we have
also used
$F(x) \sim 1/x$). Consider the second term on the r.h.s.. Since $\tilde{v}_c(k)
=\pi /|k|$, non-analytic behaviour results from both the leading non-analytic
and
leading analytic term in the expansion of $\tilde{C}(k)$:
$$
\tilde{C}(k) \sim |k|^{\Gamma - 1} + k^2
\eqno (3.41)
$$
This term is of a lower order in $k$ than the previous two considered and so
must be balanced by the 3 body term:
$$
\int_{-\infty}^\infty dx_3 \,
{\partial v(x_3) \over \partial x_3}
\left (\tilde{\rho}_{s_1s_2+}^T
(k,0,x_3) - \tilde{\rho}_{s_1s_2-}^T(k,0,x_3) \right )
\sim {s_1 \over 2} \rho i \pi {\rm sgn}(k) \Big ( |k|^{\Gamma - 1} + k^2 \Big )
\eqno (3.42)
$$
Note that for $\Gamma > 3$ the $O(k^2)$ term is dominant. Taking the inverse
transform
suggests that for large-$|x_1|$ and fixed $x_3$
$$
{\rho}_{s_1s_2+}^T
(x_1,0,x_3) - {\rho}_{s_1s_2-}^T(x_1,0,x_3) \sim
\cases{O\left ( {{\rm sgn}(x_1) \over |x_1|^\Gamma} \right ),&$2<\Gamma<3$\cr
O\left ({1 \over x_1^3} \right ),&$\Gamma >3$\cr}
\eqno (3.43)
$$
At $\Gamma = 3$, an extra logarithmic factor is expected.

The non-perturbative result (3.43) is in agreement with a low fugacity
analysis. In
particular, the $O(1/x_1^3)$ behaviour of ${\rho}_{s-+}^T
(x_1,0,x_3)$ at $O(\zeta^4)$, which is dominant for $\Gamma > 3$, originates
from the
configuration in which the mobile particle of charge $-s$ forms a dipole with
the root
charge at $x_1$. Explicitly, the first term of the multipole expansion of this
charge  configuration which gives a non-zero contribution to the cluster
integral is
$$
(u-x_1)^2(x_3-x_2){\partial^2 \over \partial x_1^2}{\partial \over \partial
x_2}
\left ( - \log |x_1 - x_2| \right ) \sim {1 \over x_1^3}
\eqno (3.44)
$$

For the 4 body correlations, an analysis of the BGY equation for the 3 body
correlations analogous to that given above leads to the conclusion that for all
$\Gamma > 2$
$$
\rho_{s_1(-s_1)s_2(-s_2)}^T(x_1,x_2,x_3,x_4) \sim {1 \over (x_1 - x_3)^2}
\eqno (3.45)
$$
as $|x_1 - x_3| \rightarrow \infty$ with $|x_1 - x_2|$ and $|x_3 - x_4|$ fixed.

\vspace{.5cm}
\noindent
{\bf 3.4.1 Exact decay of a three particle correlation at $\Gamma = 4$}

\noindent
A version of the two-component log-gas system without charge ordering, in which
the two species of charges are confined to separate, interlacing sublattices
(lattice spacing $\tau$),
is exactly solvable at the isotherm $\Gamma = 4^{(10)}$. In particular, the
exact
expression for the three particle correlation involving opposite charges is
$^{(22)}$
\renewcommand{\theequation}{3.46}
\begin{eqnarray}
\lefteqn{\rho_{+--}^T(n,m_1-1/2,m_2-1/2)}\nonumber \\
& & = {1 \over 2}{\rm Tr} \left [
B_{+-}(n,m_1-1/2)B_{--}(m_1-1/2,m_2-1/2)B_{-+}(m_2-1/2,n) \right ] \nonumber \\
& & + {1 \over 2}{\rm Tr} \left [
B_{+-}(n,m_2-1/2)B_{--}(m_2-1/2,m_1-1/2)B_{-+}(m_1-1/2,n) \right ]
\end{eqnarray}
where
$$
B_{+-}(n,m_1-1/2) = \left ( \begin{array}{cc}
\beta_{13}(l_1) & \beta_{14}(l_1) \\
\beta_{14}(l_1) & \beta_{31}(l_1) \end{array} \right ), \: \:
B_{-+}(m_2-1/2,n) = \left ( \begin{array}{cc}
\beta_{31}(l_2) & \beta_{14}(l_2) \\
\beta_{14}(l_2) & \beta_{13}(l_2) \end{array} \right )
$$
with
$$
\beta_{13}(l) = {\pi \zeta i \over \tau} (1 + \xi) \int_0^1 dt \,
{e^{2 \pi i l t} \over D(t) }
$$
$$
\beta_{14}(l) = {2\pi^2 \zeta  \over \tau^2} (1 - \xi) \int_0^1 dt \,
{e^{2 \pi i l t} \over D(t) }(1/2 - t)
$$
$$
\beta_{31}(l) = {2\pi^3 \zeta  \over \tau^3 i} (1 + \xi) \int_0^1 dt \,
{e^{2 \pi i l t} \over D(t) }t(t - 1)
$$
$$
D(t) := 1 - 16 \xi (t^2 - t + 1/8) + \xi^2, \quad \xi :=
 \zeta^2 \pi^4 / \tau^4,
\quad l_1 := n - (m_1 - 1/2) \quad l_2:=m_2-1/2-n
$$
and
$$
B_{--}(m_1-1/2,m_2-1/2) = \left ( \begin{array}{cc}
\beta_{11}(l_3) & \beta_{21}(l_3) \\
\beta_{12}(l_3) & \beta_{11}(l_3) \end{array} \right )
$$
with
$$
\beta_{11}(l) =\int_0^1 dt \,
{e^{2 \pi i l t} \over D(t) }[\xi^2 - 4 \xi t (1-t) - 4 \xi (t-1/2)^2]
$$
$$
\beta_{12}(l) =4(\tau /\pi) \xi i\int_0^1 dt \,
{e^{2 \pi i l t} \over D(t) }(t - 1/2)
$$
$$
\beta_{21}(l) = {16 (\pi / \tau) \xi \over i}\int_0^1 dt \,
{e^{2 \pi i l t} \over D(t) }(t - 1/2)t(t-1)
$$
$$
l_3:=m_1 - m_2
$$

We seek the asymptotic behaviour of (3.46) for $|m_2| \rightarrow \infty$ with
$n$ and $m_1$ fixed. Now, straightforward integration by parts of the above
integrals gives
$$
\beta_{13}(l) \sim {1 \over l} \quad \beta_{31}(l) \sim {1 \over l^3}
\quad \beta_{14}(l) \sim {1 \over l^2}
$$
$$
\beta_{11}(l) \sim {1 \over l^2} \quad \beta_{12}(l) \sim {1 \over l}
\quad \beta_{21}(l) \sim {1 \over l^3}
$$
and thus
$$
B_{--}(m_1 - 1/2,m_2 - 1/2)B_{-+}(m_2-1/2,n) \sim \left (
\begin{array}{cc}
 {1 \over m_2^5} & {1 \over m_2^4} \\
{1 \over m_2^4} & {1 \over m_2^3}
\end{array} \right )
\eqno (3.47a)
$$
$$
B_{+-}(n,m_2 - 1/2)B_{--}(m_2-1/2,m_1-1/2) \sim \left (
\begin{array}{cc}
 {1 \over m_2^3} & {1 \over m_2^4} \\
{1 \over m_2^4} & {1 \over m_2^5}
\end{array} \right ) \quad
\eqno (3.47b)
$$
Substituting (3.47) in (3.46) gives that
$$
\rho_{+--}^T(n,m_1-1/2,m_2-1/2) \mathop{\sim}
\limits_ {m_2 \rightarrow \infty}
{1 \over m_2^3}
$$
in agreement with the prediction (3.43).

\vspace{.5cm}
\noindent
{\bf 4. SUMMARY AND COMMENTS}

\noindent
We have undertaken a systematic study of properties of the particle and charge
correlation functions in the two-dimensional Coulomb gas confined to a
one-dimensional
domain, with and without charge ordering. As a result some new properties have
emerged.

Consider first the system with charge ordering. In the high temperature phase
we
have found an $O(1/x^4)$ decay of the two-particle correlations as given by
(2.18).
The system does not screen an external charge density, as this would lead to
(1.4b) and thus an $O(1/x^2)$ decay. However the underlying reference system at
$\Gamma =0$ does screen an internal charge. This allows an expansion of the
large-$x$ two-particle correlations to made about $\Gamma =0$ and leads to the
$O(1/x^4)$ decay. For $\Gamma \rightarrow 2^+$, it was shown in subsection 2.2
that the asymptotic charge-charge correlation can be resummed to all orders in
the fugacity, and that the Kosterlitz renormalization equations result. The
radius of convergence of the quantity $\Delta$ (2.19) as given by (2.37)
gives the dependence of the critical coupling on the fugacity, analogous to the
situation with the resummed dielectric constant in the 2dCG$^{(7)}$.
Furthermore,
the configurations contributing to the asymptotic charge-charge correlation
were identified as nested dipoles, analogous to situation in the scaling
region of the Kosterlitz-Thouless transition of the 2dCG. Away from criticality
in the low temperature phase the two-particle correlations have an $O(1/r^2)$
decay as given by (2.44). In a low fugacity perturbation expansion this
behaviour can
be understood as resulting from the dipole-dipole interaction between neutral
clusters
of particles about each root particle. In all phases the correlations must obey
the sum rule (2.55) involving the dipole moment $p_{+-}(x)$ of the
charge distribution induced by two fixed opposite charges. This sum rule is
analogous
to the sum rule (1.3b) for the 2dCG, which relates the dielectric constant to
$p_{-+}$.

Now consider the system without charge ordering. In the high temperature
phase $\Gamma \le 2$ the system is conductive and the charge correlation
must exhibit the asymptotic behaviour (1.4b). The low fugacity analysis of the
two-particle correlations given in subsection 3.2 imply a $O(1/x^\Gamma)$ decay
for $2\le \Gamma < 4$ and an $O(1/x^4)$ decay for $\Gamma \ge 4$. In the
context
of the low fugacity expansion this behaviour results from configurations
consisting of clusters about the root charges which have a charge imbalance
equivalent to that of one particle dominating for $2 < \Gamma < 4$, while
neutral clusters dominate for $\Gamma > 4$.  Furthermore, for
$\Gamma \rightarrow 2^+$ the coefficients of the low fugacity expansion
of the large-$x$ two-particle correlations are all finite; there is no
dominance of nested dipole configurations which in fact cancel without
charge ordering. The finiteness of the coefficients indicates that the
phase transition occurs at $\Gamma = 2$ independent of the fugacity.
With a smoothly regularized potential, we have presented the sum rule
(3.25) involving the dipole moment $p_{+-}(x)$, and explicitly verified
it on a solvable model at the coupling $\Gamma = 2$.

As far as large-distance behaviours of the internal correlations are
concerned, the system with charge ordering is very similar to the familiar
$2D$ Coulomb gas. Indeed, the charge correlations have a ''fast'' decay
in the high temperature phase ($1/|x|^4$ in comparison to an exponential
decay in 2D), and as a density-dependent power law in the low-temperature
phase ($1/|x|^{\Gamma \Delta}$ in comparison to $1/|r|^{\Gamma / \epsilon}$
in 2D). The partial screening of a given pair (with sizes $X$ in 1D and
$R$ in 2D) by smaller pairs takes the same mathematical form for both
systems, because of the occurence of the integrals
$$
\int_{\sigma < x < X} dx \, {d \over dx} \log |x|
\eqno (4.1a)
$$
and
$$
\int_{\sigma < r < R} d\vec{r} \, \Big ( \vec{\nabla} \log r \Big )^2
\eqno (4.1b)
$$
which both diverge logarithmically when $X$ and $R$ go to infinity. The
integral (4.1a) arises from the dipole-dipole potential between the
screening pair and the root charges. The resulting screening contribution
is proportional to the average of the dipole carried by this pair, which
does not vanish by virtue of the charge ordering constraint.

Of course, the screening contribution associated with (4.1a) disappears
in the system without charge ordering. The leading contribution of a
screening pair then involves
$$
\int_{\sigma < x < X} dx \, \left ( {d \over dx} \log |x|\right )^2
\eqno (4.2)
$$
which is linked to the fluctuations of the dipole-charge potential between
this pair and the root charges. The integral (4.2) does converge in the
limit $X \rightarrow \infty$, and consequently the collective effects do not
affect the power which controls the algebraic decay of the charge
correlations. The Kosterlitz-Thouless transition temperature is then
expected to remain constant $(\Gamma = 2)$ at low densities. This is
contrary to the findings of a recent computer simulation$^{(23)}$, where
the transition coupling appeared to decrease with decreasing density.
However, we believe these findings are linked to finite-effects which
become more and more important as the density decreases. Indeed, in the
zero-density limit and with $1 < \Gamma < 2$, the proportion of free
charges goes to zero (see e.g.~ref.~[15]), a behaviour which is reminiscent of
the
short-range collapse at $\Gamma = 1$ for the system without hard-core
regularization. In this regime of parameters, the finite systems considered
in the numerical simulations are not efficient for perfectly screening
external charges because of the very small number of free charges which
are available. A similar mechanism also occurs in 2D$^{(24)}$, and even in
3D$^{(25)}$ for the restricted primitive model with $1/r$ interactions
(in that case there is no Kosterlitz-Thouless transition).
Eventually, we note that at high densities, despite difficult problems of
sampling, the numerical simulations$^{(23)}$ are compatible with the
theoretical predictions given here and in ref.~[3].

Our predictions should be valid at low densities. At higher densities,
the critical Kosterlitz-Thouless transition line in the $(\rho , T)$
plane might bifurcate, at a tricritical point, into a first-order
liquid-gas coexistence curve. In fact, for the 2D Coulomb gas, this
has been observed in computer simulations by  Caillol and Leveque$^{(26)}$.
More recently, an extended Debye-H\"uckel theory  of Levin et al$.^{(27)}$
suggests a tricritical
point at a much lower density.
To our knowledge, such first-order transitions have not yet been observed for
the present models in 1D. At a theoretical level, their study is beyond
the scope  and the methods of this paper. In particular, a correct
description of eventual oscillatory correlations$^{(28)}$ beyond a
Fisher-Widom line requires further resummations of the low fugacity
expansions combined to a detailed account of the short-range effects
which depend on the hard-core regularization.
At the moment, note that for the model without charge ordering, the
exact results for the solvable isotherm $\Gamma = 2^{(10,11)}$ seem
to exclude the above bifurcation process.

\vspace{.5cm}

\noindent
{\bf Acknowledgements}

\noindent
We thank the referees for their considered comments. P.J.F. was supported by
the Australian Research Council.

\pagebreak
\noindent
{\bf Appendix A}

\vspace{.5cm}
\noindent
Our objective in this appendix is to show how the integrals in (2.23)
can be analysed to determine the asymptotic charge density $C_\Delta^{(4)}
(r)$. We recall $C_\Delta^{(4)}(r)$ is defined as the portion of $C(r)$
proportional
to $\zeta^4$ which contributes to  the correct leading order behaviour of
(2.19) in the
limit $\Gamma \rightarrow 2^+$.

We will illustrate our method by calculating the contribution to
$C_\Delta^{(4)}(r)$ from $\rho_{++}^{(4)}(r)$. Scaling the integration
variables by
$$
y_1 \mapsto ry_1 \qquad {\rm and} \qquad y_2 \mapsto ry_2
$$
we read off from (2.23a) that
$$
\rho_{++}^{T(4)}(r) = { \zeta^4 \over r^{2\Gamma - 2} } \left [
\int_{\sigma /r}^{1 - \sigma /r} dy_1 \int_{1 + \sigma /r}^\infty dy_2 \,
 \left ( { y_2 - y_1 \over (y_2 - 1)y_2(1-y_1)y_1 } \right )^\Gamma
 - {r^{2\Gamma - 2} \over (\Gamma - 1)^2} \right ]
\eqno ({\rm A}1)
$$
The fixed root points are now at 0 and 1. Let $\mu$ be a positive constant
such that $\sigma /r < \mu \ll  1$. Then for the region of integration
$\mu < y_1 < 1-\mu$ and $y_2 > 1 + \mu$ in (A1) the integrand is bounded
uniformly in $r$ and is integrable. Hence the contribution to the double
integral is $O(1)$, and from (2.19) the corresponding contribution to
${\Delta}^{(4)}$ is $O( \zeta^4 / (\Gamma - 2))$.
To analyse the contribution from the remaining region of integration we
decompose it into the subregions indicated by the particle configurations in
Fig. 2.

Consider subregion A in which
$$
\sigma /r \le y_1 \le \mu \qquad {\rm and} \qquad \sigma / r \le y_2 - 1 \le
\mu
$$
Since $y_1$ and $y_2 - 1$ are small variables we can expand the integrand:
$$
\left ( {y_2 - y_1 \over (y_2 - 1)y_2(1 - y_1)y_1} \right )^\Gamma
\: \sim \:
{1 \over (y_2 - 1)^\Gamma y_1^\Gamma }
(1 + \Gamma y_1(y_2-1)+ \dots )
$$
This shows that the leading large-$r$ contribution to the double integral from
subregion A is
$$
{r^{2\Gamma - 2} \over (\Gamma - 1)^2} + { \Gamma \over (\Gamma - 2)^2 }
( r^{\Gamma - 2} - 1)^2
\eqno ({\rm A}2)
$$
Substituting (A1) in (A2)  shows that the first term cancels. To leading order
for large-$r$ the second term gives a contribution
$$
\rho_{++}^{T(4)}(r) \: \sim \: {\zeta^4 \Gamma \over (\Gamma - 2)^2 r^2}
$$
The first moment of this term (considered as a function defined on $[\sigma
,\infty)$)is infinite. However the entire second term in (A2) is exactly
cancelled in
the formula (2.24) for $C(r)$ by an identical term contributing to
$\rho_{+-}^{(4)T}(r)$, so there is no net contribution to $\Delta^{(4)}$.

Next consider the subregion B(i) in which
$$
\sigma /r \le 1 - y_1 < \mu  \qquad {\rm and } \qquad
\sigma /r \le y_2 - 1 \le 1 - y_1
\eqno ({\rm A}3)
$$
Expanding the integrand in (A1) given these conditions we have
$$
\left ( {y_2 - y_1 \over (y_2 - 1)y_2(1-y_1)y_1} \right )^\Gamma
\: \sim \: { 1 \over (1 - y_1)^\Gamma} \left ( 1 +  \Gamma { 1 - y_1 \over
y_2 - 1} + \dots  \right )
$$
The corresponding contribution to the double integral is
$$
\int_{1-\mu}^{1-\sigma /r} {dy_1 \over (1-y_1)^\Gamma }
\int_{1+\sigma/r}^{1+(1-y_1)}dy_2
+\Gamma \int_{1-\mu}^{1-\sigma /r}{dy_1 \over (1-y_1)^{\Gamma -1}}
\int_{1+\sigma/r}^{1+(1-y_1)}{dy_2 \over y_2 - 1}
\eqno ({\rm A}4)
$$
When evaluated for large-$r$ these two integrals give
$$
{ (r /\sigma)^{\Gamma - 2} \over \Gamma - 2 }
\eqno ({\rm A}5a)
$$
and
$$
\Gamma \left ( -{1 \over (\Gamma - 2)^2} \left [ \left ({r \over \sigma }
\right )^{\Gamma - 2}
- 1 \right ] + { 1 \over \Gamma - 2 }   \left ( {r \over
\sigma}\right )^{\Gamma - 2} \log r \right )
\eqno ({\rm A}5b)
$$
respectively.
The corresponding contribution to $\Delta^{(4)}$ from (A5a) is
$O(1/(\Gamma - 2)^2)$ while the contribution from (A5b) is
$O(1/(\Gamma - 2)^3)$. Hence (A5a) does not contribute to $C_\Delta^{(4)}(r)$.

The analysis for region B(ii) proceeds analogously to that of region B(i)
just presented. We again find contributions (A5a) and (A5b). In regions C(i)
and C(ii)
we find a contribution to $\Delta^{(4)}$ which is $O(1/(\Gamma - 2)^2)$.
This region therefore does not contribute to $C_\Delta^{(4)}(r)$, and since all
regions are now exhausted, the terms in the asymptotic expansion of
$\rho_{++}^{T(4)}(r)$ which contribute to $\Delta^{(4)}$ give a contribution
to the latter which is $O(1/(\Gamma - 2)^3)$. Adding all such terms,
including the term corresponding to (A2) which gives a divergent
contribution to $\Delta^{(4)}$, we obtain the expansion

$$
\rho_{++}^{T(4)}(r) \: \sim \: \zeta^4 \left [
{ \Gamma \over (\Gamma - 2)^2} {1 \over r^{2 \Gamma - 2}} \Big ( 1 - \left
( {r \over \sigma }\right )^
{\Gamma - 2} \Big )^2 \right. \hspace{4cm}
$$
$$
+ \left. { 2\Gamma \over r^{2\Gamma - 2}} \left (-{1 \over (\Gamma - 2)^2}
\Big (\left ( {r \over \sigma} \right )^{\Gamma - 2} - 1 \Big ) + {1 \over
\Gamma - 2} \left( {r \over \sigma} \right)^{\Gamma - 2}
\log r \right ) \right ]
\eqno ({\rm A}6a)
$$
The first term originates from subregion A while the remaining two terms
have equal contribution from subregions B(i) and B(ii).

A similar approach suffices to analyse the terms  in
the asymptotic expansion of (2.23c) and (2.23b) which contribute to
$\Delta^{(4)}$.
We find
$$
\rho_{-+}^{T(4)}(r) \: \sim \:
{ \zeta^4 \Gamma \over (\Gamma - 2)^2 r^{2 \Gamma - 2}}
\left ( \left ({r \over \sigma} \right )^{\Gamma - 2} - 1\right )^2
\eqno ({\rm A}6b)
$$
and
$$
\rho_{+-}^{T(4)}(r) \: \sim \: \zeta^4 \left [
{ \Gamma \over (\Gamma - 2)^2 }{1 \over r^{2 \Gamma - 2}} \Big ( 1 - \left ({r
\over \sigma} \right )^
{\Gamma - 2}\Big )^2\right.\hspace{4cm}
$$
$$
+ \left. { 8\Gamma \over r^{2\Gamma - 2}} \left (-{1 \over (\Gamma - 2)^2}
(\left ({r \over \sigma} \right )^{\Gamma - 2} - 1) + {1 \over \Gamma - 2}
\left
({r \over \sigma}\right )^{\Gamma - 2}
\log r \right ) \right ]
\eqno ({\rm A}6c)
$$
The particle configurations corresponding to the subregions of integration
contributing to the terms in these expansions  are given
in Fig. 3.

Forming the combination (2.24) of (A6a),(A6b) and (A6c) required to form
$C(r)$ we see that a lot of cancellation takes place, and the result (2.27)
of the text results. Furthermore, it is important to observe that due to the
cancellations only $\rho_{+-}^{T(4)}(r)$ contributes to $C_\Delta^{(4)}$ and
this contribution is restricted to the region of integration depicted in
b(ii) of Fig. 3. Analogous to (A5b) this contribution can be written as
a double integral which is given by (2.28) of the text. The integrand of the
double integral is obtained from the
expansion of the integrand (excluding the factor of $1 / r^ \Gamma $) in the
second double integral of (2.23b):
$$
\left ( { x (y - r) \over (y - x) y ( x - r)} \right )^ \Gamma -
\left ( {1 \over y - x} \right )^\Gamma
\: \sim \: {\Gamma \over (x-r) (y - x)^{\Gamma -1}}
\eqno ({\rm A}7)
$$
valid for $x-r$ and $y-x$ small and positive with $y - x<x-r$. The domain of
integration is that implied in configuration b(ii) of Fig. 3.

\vspace{1cm}
\noindent
{\bf APPENDIX B}

\vspace{.5cm}
\noindent
Our objective here is to study the leading large-$r$ behaviour of the
formula (3.4) for $C^{(4)}(r)$ as a function of $\Gamma$. We begin by changing
variables $x_1 \mapsto ru \: x_2 \mapsto rv$, so that (3.4) reads
$$
C^{(4)}(r) = {q^2 \zeta^4 \over r^{2 \Gamma - 2}} \int^* du dv
\left ( \left | {u-v \over (v-1)(u-1)uv} \right |^\Gamma
-2 {\rm S} \left | {u(v-1) \over (u-v)(u-1)v} \right |^\Gamma
-{2 \over |u-v|^\Gamma} \right )
\eqno ({\rm B}1)
$$
where * denotes the integration region
$$
\{ (u,v)\in {\bf {\rm R}}^2: \: |u-v| > \sigma /r, \: |u|,|v| > \sigma /r,
\: |u-1|,|v-1| >
\sigma /r \}
$$
Due to the  integrand being symmetric in $u$ and $v$, and unchanged by the
mappings $u \mapsto 1-u, \: v \mapsto 1-v $, we can restrict the integration
region to
$$
\{ (u,v)\in {\bf {\rm R}}^2: \: u>v, \, v>1-u, \, |u-v| > \sigma /r, \:
|u-1|,|v-1| > \sigma /r \}
\eqno ({\rm B}2)
$$
provided we multiply the integral by 4. Let us divide the integration region
(B2) into labelled and unlabelled regions as indicated in Figure 4.

All regions outside those labelled in Fig. 4  give a contribution to
$C^{(4)}(r)$ which is $O(1/r^{2\Gamma - 2})$. This follows from (B1) since
the
range of the integrand is independent of $r$ in these regions. The range
of the integrand is not independent of $r$ in the labelled regions of
Fig. 4, and consequently further calculations are needed.

Let us illustrate our method of analysing the leading order contribution
from the labelled regions by considering region D. The total contribution
to $C^{(4)}(r)$ from this region is given by
$$
{q^2 \zeta^4 \over r^{2\Gamma - 2}} \int_{1+\mu}^\infty du \,\left (
\int_{\sigma /r}
^\mu + \int_{-\mu}^{-\sigma /r} \right ) dv \, F(u,v)
\eqno ({\rm B}3)
$$
where $F(u,v)$ is the integrand in (B1).
For large-$r$ the leading contribution to (B3) comes from the neighbourhood
of $|v| = \sigma /r$. We can thus ignore the last term in (B1), and the term
implicit in the symmetrized form of the second term in (B1). Expanding the
remaining two terms for small-$v$ gives
$$
{2 \Gamma \over |v|^\Gamma (u-1)^\Gamma} \left ( -{v \over u} + v
-{1 \over 2} \left ({v \over u} \right )^2 + {v^2 \over 2} + O(v^3) \right )
$$
in place of $F(u,v)$ in (B3).

Evaluating this integral to leading order for large-$r$ gives behaviours
$$
O\left ( {1 \over r^{2 \Gamma - 2} }\right ) \: 2 < \Gamma <3, \quad
O\left ( {\log r \over r^4 }\right ) \: \Gamma = 3, \quad
O\left ( {1 \over ( \Gamma - 3) r^{\Gamma + 1}} \right ) \: \Gamma > 3
\eqno ({\rm B}4)
$$
Proceeding similarly, we find the leading large-$r$ contributions from the
labelled regions of Fig. 4 to be as in the table below.\\[.5cm]

\begin{tabular}{cccc}
Region & \multicolumn {3}{c}{Contribution to $C^{(4)}(r)$}\\[.3cm]
{}& $2< \Gamma < 3$ & $\Gamma = 3$ & $\Gamma > 3$\\[.3cm]
A,B,C,D & O$\left( {1 \over r^{2 \Gamma - 2}} \right )$ &
O$\left ( {\log r \over r^4 }\right )$ & O$\left ({1 \over (\Gamma - 3)r^{
\Gamma + 1} } \right ) $\\[.3cm]
$\alpha $ & O$\left ( {1 \over r^\Gamma} \right )$ & O$\left ({1 \over r^3}
\right )$ & O$\left ( {1 \over r^\Gamma} \right )$\\[.3cm]
$\beta $ & O$\left( {1 \over r^{2 \Gamma - 2}} \right )$ &
O$\left ( {\log^2 r \over r^4 }\right )$ & O$\left ({1 \over (\Gamma - 3)^2r^4}
\right )$
\end{tabular}

\vspace{.5cm}

Hence the leading order contribution to $C^{(4)}(r)$ comes from region
$\alpha$ for $2 < \Gamma < 4$ and from region $\beta$ for $\Gamma > 4$.
These regions correspond to the charge configurations of Fig. 5.

\pagebreak

\noindent
{\bf Figure 1.} The two distinct nested dipoles at $O(\zeta^6)$. The square
braces indicate a screening operator which connects the charge on the left end
with
the dipole on the right end.
\medskip

\noindent
{\bf Figure 2.} Some regions of integration in the  integral
(A1) for $\rho_{++}^{T(4)}(r)$.
\medskip

\noindent
{\bf Figure 3.}  The configuration (a) gives the term in (A6b), while the
configuration b(i) gives the first term in (A6c). The configurations
b(ii) and b(iii) both contribute equally to the remaining terms in
(A6c). These configurations each have a multiplicity factor of 4, which
corresponds to the four ways of placing the mobile pair x and y on either side
of the two root charges.
\medskip

\noindent
{\bf Figure 4.} Regions of integration in (B2) which are used to analyse the
integral (B1) for $C^{(4)}(r)$. A distance $\sigma /r$ either side of each
dotted line bordering or contained within a region is to be excluded.
\medskip

\noindent
{\bf Figure 5. } Charge configurations corresponding to regions $\alpha$ and
$\beta$ in Figure 4, which give the leading order contribution to $C^{(4)}(r)$
for $2<\Gamma<4$ and $\Gamma >4$ respectively. The arrows indicate
alternative placement of mobile charges.

\pagebreak

\noindent
{\bf References}

\noindent
1. P.W. Anderson, G. Yuval and D.R. Hamman, Phys. Rev. B {\bf 1}: 4464 (1970)\\
[.1cm]
2. A. Schmid, Phys. Rev. Lett. {\bf 51}: 1506 (1983) \\[.1cm]
3. H. Schultz, J. Phys. A {\bf 14}: 3277 (1981) \\[.1cm]
4. J. M. Kosterlitz, J. Phys. C {\bf 7}: 1046 (1974) \\[.1cm]
5. P. Minnhagen, Rev. Mod. Phys. {\bf 59}: 1001 (1987) \\[.1cm]
6. K. D. Schotte and U. Schotte, Phys. Rev. B {\bf 4}: 2228 (1971) \\[.1cm]
7. A. Alastuey and F. Cornu, J. Stat. Phys. {\bf 66 }: 165 (1992) \\[.1cm]
8. P.J. Forrester, J. Stat. Phys. {\bf 42}: 871 (1986) \\[.1cm]
9. P.J. Forrester, J. Stat. Phys. {\bf 45}: 153 (1986) \\[.1cm]
10. P.J. Forrester, J. Stat. Phys. {\bf 59}: 57 (1989) \\[.1cm]
11. M. Gaudin, J. Phys. (Paris) {\bf 46}: 1027 (1985) \\[.1cm]
12. E. Meeron, J. Chem. Phys. {\bf 28}: 630 (1958); R. Abe, Prog. Theor.
Phys. {\bf 22}: 213 (1959) \\[.1cm]
13. A. Alastuey and Ph. A. Martin, Phys. Rev. A {\bf 40}: 6485 (1989) \\[.1cm]
14. E. R. Speer, J. Stat. Phys. {\bf 42}: 895 (1986) \\[.1cm]
15. P. J. Forrester and B. Jancovici, J. Stat. Phys. {\bf 69}: 163 (1992)
\\[.1cm]
16. J. M. Kosterlitz and D. J. Thouless. J. Phys. C {\bf 6}: 1181
(1973)\\[.1cm]
17. Ph. A. Martin, Rev. Mod. Phys. {\bf 60}: 1075 (1988)\\[.1cm]
18. B. Jancovici, J. Stat. Phys. {\bf 80} (July 1995)\\[.1cm]
19. A. Alastuey, F. Cornu and A. Perez, Phys. Rev. E {\bf 49}: 1077
(1994)\\[.1cm]
20. P.J. Forrester, J. Stat. Phys. {\bf 51}: 457 (1988)\\[.1cm]
21. F. Cornu and B. Jancovici, J. Chem. Phys. {\bf 90}: 2444 (1989)\\[.1cm]
22. M.L. Rosinberg, unpublished \\[.1cm]
23. G. Manificat and J.M. Caillol, J. Stat. Phys. {\bf 74}: 1309 (1994)
24. A. Alastuey, F. Cornu and B. Jancovici, Phys. Rev. A {\bf 38}:
4916 (1988) \\[.1cm]
25. J.M. Caillol, to be published in J. Chem. Phys. \\[.1cm]
26. J.M. Caillol and D. Levesque, Phys. Rev. B {\bf 33}: 499 (1986)\\[.1cm]
27. Y. Levin, X. Li and M.E. Fisher, Phys. Rev. Lett. {\bf 73}: 2716
(1994)\\[.1cm]
28. R. Evans and L. de Carvalho, to be published in Mol. Phys.

\end{document}